\documentclass[epj,final]{svjour}

\usepackage{times}
\usepackage{graphicx}
\usepackage{amsmath}
\usepackage{enumerate}
\usepackage{indentfirst}
\usepackage{xspace}

\newcommand{\numu}{\ensuremath{\nu_{\mu}}\xspace}
\newcommand{\nue}{\ensuremath{\nu_{e}}\xspace}

\newcommand{\piplus}{\ensuremath{\pi^{+}}\xspace}
\newcommand{\pip}{\ensuremath{\pi^{+}}\xspace}

\newcommand{\pim}{\ensuremath{\pi^{-}}\xspace}

\newcommand{\kam}{\ensuremath{\mbox{K}^{-}}\xspace}
\newcommand{\prot}{\ensuremath{\mbox{p}}\xspace}

\newcommand{\ddpiplus}{\ensuremath{d^2\sigma^{\pi^{+}}/dpd\Omega}\xspace}

\newcommand{\rad}{\ensuremath{\mbox{rad}}\xspace}
\newcommand{\GeV}{\ensuremath{\mbox{GeV}}\xspace}
\newcommand{\MeV}{\ensuremath{\mbox{MeV}}\xspace}
\newcommand{\GeVc}{\ensuremath{\mbox{GeV}/c}\xspace}

\newcommand{\cmcube}{\ensuremath{\mbox{cm}^3}\xspace}

\newcommand{\cm}{\ensuremath{\mbox{cm}}\xspace}
\newcommand{\mm}{\ensuremath{\mbox{mm}}\xspace}

\newcommand{\mrad}{\ensuremath{\mbox{mrad}}\xspace}

\newcommand{\m}{\ensuremath{\mbox{m}}\xspace}

\newcommand{\ches}{{S_{\mathrm{c}}}}
\newcommand{\pr}{\mathrm{p}}

\begin{document}

\title{Measurement of the production cross-section of positive pions in the collision of 8.9 GeV/c protons on beryllium}
\titlerunning{$\pi^{+}$ production cross-section in the collision of 8.9 GeV/c protons on beryllium}

\author{
  M.G.~Catanesi\inst{1} \and 
  E.~Radicioni\inst{1} \and
  R.~Edgecock\inst{2} \and
  M.~Ellis\inst{2}\thanks{Now at FNAL, Batavia, Illinois, USA.} \and
  S.~Robbins\inst{2}\thanks{Jointly appointed by Nuclear and Astrophysics Laboratory, University of Oxford, UK.}
  \fnmsep \thanks{Now at Codian Ltd., Langley, Slough, UK.} \and
  F.J.P.~Soler\inst{2}\thanks{Now at University of Glasgow, UK.} \and
  C.~G\"{o}\ss ling\inst{3} \and
  S.~Bunyatov\inst{4} \and 
  G.~Chelkov\inst{4} \and     
  D.~Dedovitch\inst{4}
% ...
}

\institute{
  Universit\`{a} degli Studi e Sezione INFN, Bari, Italy \and 
  Rutherford Appleton Laboratory, Chilton, Didcot, UK \and
  Institut f\"{u}r Physik, Universit\"{a}t Dortmund, Germany \and
  Joint Institute for Nuclear Research, JINR Dubna, Russia
% ...
}

\date{Received: date / Revised version: date}

\abstract{The double-differential production cross-section of positive pions, \ddpiplus, 
measured in the HARP experiment is presented.  The incident particles are 8.9 \GeVc
protons directed onto a beryllium target with a thickness of  5\% of a nuclear interaction length.  
The measured cross-section has a direct impact on the prediction of neutrino fluxes for the 
MiniBooNE and SciBooNE experiments at Fermilab.  After cuts, 13 million protons on target
produced about 96,000 reconstructed secondary tracks which were used in this analysis.  
Cross-section results are presented in the kinematic range 0.75~\GeVc  $\leq p_{\pi} \leq$ 6.5~\GeVc 
and 30~mrad $\leq \theta_{\pi} \leq$ 210~mrad in the laboratory frame. 
\PACS{
      {PACS-key}{discribing text of that key}   \and
      {PACS-key}{discribing text of that key}
     } 
} 
\maketitle

\section{Introduction}

The HARP experiment was designed to make measurements of hadron yields from a large range
of nuclear targets and for incident particle momenta from 1.5~\GeVc\ -- 15 \GeVc.
Among its primary goals were to contribute to the 
detailed understanding of neutrino beams of several experiments, 
including: 
\begin{itemize}
\item The K2K experiment, which has recently published its final results \cite{ref:k2kfinal}
confirming the evidence of atmospheric oscillations observed by Super-Kamiokande \cite{ref:superKatmoOsc}.
\item The MiniBooNE experiment \cite{ref:minibooneProposal}, which recently excluded \cite{ref:minibooneOsc} 
two neutrino appearance-only oscillations as an explanation of the LSND anomaly \cite{ref:LSND}, 
in the hypothesis that the oscillations of neutrinos and antineutrinos are the same.
The MiniBooNE detector
will also be used to measure neutrino interaction cross-sections for which an absolute prediction of
neutrino fluxes becomes of particular importance.
\item The SciBooNE experiment \cite{ref:sciboone}, which will take data in the same neutrino beam used by MiniBooNE 
in order to perform a precision measurement of neutrino cross-sections in the energy region
around 1 \GeV.
\end{itemize}

The calculation of the flux and relative neutrino composition of a neutrino beam requires
a precise measurement of the interaction cross-section between the beam particles
and the target material. In the case of the
K2K and the MiniBooNE and SciBooNE experiments, the dominant
component of the beam (muon neutrinos) comes from the decay of positive pions produced in
the collisions of incident protons on a nuclear target. To compute the $\nu_\mu$ flux
one needs a $4 \pi$ parameterization of the differential cross section, \ddpiplus, which,
in order to be reliable, must be based on a wide-acceptance, precise measurement. 
The physics program of the HARP experiment includes the measurement of these cross-sections.

An earlier publication reported measurements of the
\piplus\ cross-sections from an aluminum target 
at 12.9 \GeVc\ \cite{ref:alPaper}. This corresponds to the energies of the KEK PS
and the target material used by the K2K experiment.  
The K2K oscillation result relies on both the measurement of an overall deficit of 
muon neutrino interactions and on the measurement of an energy spectrum deformation observed at 
the Super-Kamiokande far detector compared to the no-oscillations expectations. Introducing the
HARP experimental input in the K2K oscillation analysis has been particularly beneficial in reducing 
the systematic uncertainty in the overall number of muon neutrino interactions expected;
the near-to-far flux extrapolation contribution to this uncertainty was reduced from 5.1\% \cite{ref:k2koriginal} 
to 2.9\% \cite{ref:k2kfinal}. 

Our next goal is to contribute to the understanding of the MiniBooNE and SciBooNE neutrino fluxes. 
They are both produced by the Booster Neutrino Beam at 
Fermilab which originates from protons accelerated to 8.9 \GeVc\ by 
the Fermilab Booster before being collided against a beryllium target. As was the case for the K2K
beam, an important input for the calculation of the resulting $\nu_\mu$ flux is the
\piplus\ production cross-sections from a beryllium target 
at 8.9 \GeVc, which will be presented in this paper.

\begin{figure*}
  \begin{center}
    \includegraphics[width=12cm]{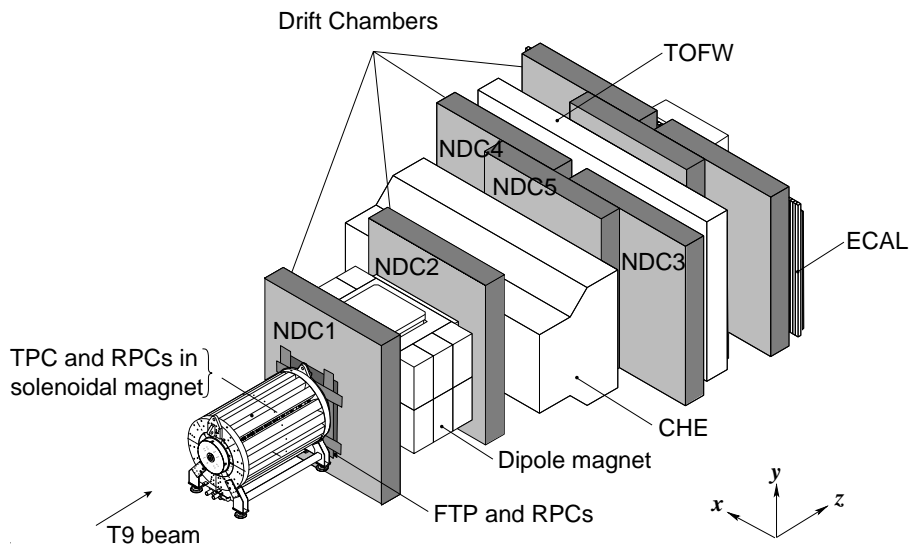}
    \caption{\label{fig:harpDet} Schematic layout of the HARP spectrometer.
      The convention for the coordinate system is shown in the lower-right
      corner. The three most downstream (unlabeled) drift chamber modules are only partly
      equipped with electronics and not used for tracking.}
  \end{center}
\end{figure*}

The HARP experimental apparatus is effectively divided into two tracking and particle identification 
sub-systems, a small-angle/high-momentum detection system ($\theta$: 0--0.25 rad, $p$: 0.5--8 \GeVc) 
and a large-angle/low-momentum system ($\theta$: 0.35--2.15 rad, $p$: 0.1--0.8 \GeVc). 
Fig. \ref{fig:harpDet} shows a schematic of the
HARP detector.  Five modules of the NOMAD drift chambers \cite{ref:NOMAD_NIM_DC} (NDC1-5) 
and the dipole magnet comprise the forward spectrometer; 
a time-of-flight wall (TOFW), Cherenkov detector (CHE) and electromagnetic calorimeter (ECAL) make 
up the particle identification (PID) system.
The large angle tracking and PID system is comprised of a time projection chamber (TPC) and resistive plate chambers (RPCs).  
The relevant meson production for the creation of the MiniBooNE and SciBooNE
neutrino fluxes is forward (0--0.30 rad) and at large momenta (0.5--6 \GeVc). These ranges are best covered
by the forward tracking system and PID detectors and so the large angle system is not used in the present analysis. 

The results reported here are based on data taken in 2002 in the T9 beam of the CERN PS.  About 2.3 
million incoming protons were selected.  After cuts, 95,897 reconstructed secondary tracks were used
in the analysis. The absolute normalization of the cross-section was determined using 204,295 `minimum-bias'
trigger events.

The analysis used in the calculation of the p-Be \piplus production cross-sections being presented here follows largely from
that used in a previous publication of p-Al \piplus production cross-sections \cite{ref:alPaper}. 
The present analysis description, therefore, will focus on the \emph{differences} in the analysis compared to the p-Al publication. 
 
This paper is organized as follows. In Section \ref{sec:diff} we summarize the main changes made to the analysis
since the p-Al publication. 
In Section \ref{sec:xsec} we describe the calculation of the cross-section and define its components. 
The following three sections expand on aspects of the analysis where significant changes have been made since the
previous publication.
Section \ref{sec:analysistracks} describes event and track selection and reconstruction efficiencies.
Section \ref{sec:momentum} discusses the determination of the momentum resolution and scale in the forward spectrometer.
Section \ref{sec:pid} summarizes the particle identification techniques.
Physics results are presented in Section \ref{sec:results}.
Section \ref{sec:miniboone} discusses the relevance of these results to neutrino experiments.
Finally, a summary is presented in Section \ref{sec:conclusions}.

\section{Summary of analysis changes since the HARP p-Al publication}\label{sec:diff}

The analyses of the 12.9 \GeVc p-Al data and the 8.9 \GeVc p-Be data are largely the same.  
To avoid repetition of information, the reader is referred to that earlier publication for many details
not directly discussed in the present paper.  The sections concerning the experimental apparatus,
the description of the tracking algorithm for the forward spectrometer
and the method of calculating the track reconstruction efficiency are all directly valid here.
The method of particle identification has not changed; it is only the PID detector hit selections and therefore their
response functions which have been significantly improved.  
The most important improvements introduced in this analysis compared with the one presented in \cite{ref:alPaper} are:

\begin{itemize}
\item An improvement in the $\chi^2$ minimization performed as part of the 
tracking algorithm has eliminated the anomalous dip in tracking efficiency above
4 \GeVc shown in \cite{ref:alPaper}. The tracking efficiency is now $\geq$ 97\% everywhere above 2 \GeVc. 
(See Sec. \ref{sec:reconEff}).   
\item Studies of HARP data other than that described here have enabled a validation of our Monte Carlo simulation
of low-energy hadronic interactions in carbon.  Specifically, we have compared low energy p+C and $\pi$+C cross-sections
to distributions from the Binary cascade \cite{ref:binaryCascade} and Bertini intra-nuclear cascade 
\cite{ref:bertini} hadronic interaction models used to simulate the secondary interactions of
p, n and $\pi^{\pm}$.  The material in the HARP forward spectrometer where tertiary tracks might be produced
is predominantly carbon.  Consequently, the systematic error on the subtraction of tertiary
tracks has been reduced from 100\% in \cite{ref:alPaper} to 50\%. (See Sec. \ref{sec:absorption}). 
\item Analysis techniques were developed for comparing the momentum reconstructions in data and
Monte Carlo allowing data to be used to fine-tune the drift chamber simulation parameters.  
These efforts have reduced the momentum scale uncertainty 
from 5\% in \cite{ref:alPaper} to 2\% in the present analysis and provided a better understanding of
the momentum smearing caused by the HARP spectrometer, including our knowledge 
of the non-Gaussian contributions to the resolution function. (See Sec. \ref{sec:momentum}). 
\item New selection cuts for PID hits in TOFW and in CHE have resulted in much reduced backgrounds and negligible
efficiency losses.  Consequently, the uncertainty on the cross-section arising from particle identification was reduced 
by a factor of seven to 0.5\% making PID 
now a negligible contribution to the systematic error in pion yield measurements at forward angles. (See Sec. \ref{sec:pid}).
\item Improved knowledge of the proton beam targeting efficiency and of fully correlated contributions
to track reconstruction and particle identification efficiencies have reduced the overall normalization 
uncertainty on the pion cross-section measurement from 4\% to 2\%.
\item Significant increases in Monte Carlo production  
have reduced uncertainties from Monte Carlo statistics and allowed
      studies to reduce certain systematics to be made.
\end{itemize}

The statistical precision of the data, however, is noticeably worse.
The 8.9 \GeVc beryllium and empty-target data sets are both smaller than the corresponding 12.9 \GeVc 
sets, with 73\% and 42\% of the protons-on-target for the target and empty-target configurations, respectively. 
The statistics of the target sample is further reduced by the p-Be total interaction cross-section being roughly 50\%
of the p-Al total cross-section.  

In the present paper the p-Be cross-sections are presented in 13 momentum bins from 0.75 \GeVc -- 6.5 \GeVc whereas
the p-Al cross-sections were presented in 8 bins.  This new binning was selected to attain roughly equal statistical and
systematic uncertainties - on average 6.3\% statistical and 7.4\% systematic in the 78 ($p,\theta$) bins 
- while maximizing the amount of spectral information provided by the measurement.       
It should be noted that the magnitude of fractional systematic errors arising from the momentum resolution and scale 
will be affected by the fineness of the binning.  In particular, in the present paper, the momentum scale uncertainty 
has been reduced from 5\% to 2\% since the p-Al publication yet this does not lead to a smaller systematic contribution 
on the measured cross-section.  This is expected since, simultaneous to the improved reconstruction,
most momentum bins have been narrowed by a factor of 2.  

In the end, the statistical plus systematic uncertainty on the total integrated
cross-section has improved from 5.8\% in p-Al to 4.9\% in p-Be.
Due to rebinning and larger statistical errors, the average bin-to-bin uncertainty on the differential 
cross-section has changed from 8.2\% in p-Al to 9.8\% in p-Be.          

We point out that a re-analysis of the proton-aluminum data incorporating these changes yields results consistent 
with those published in \cite{ref:alPaper} within the systematic errors reported there.


\section{Calculation of the double-differential inelastic cross-section}\label{sec:xsec}

The goal of this analysis is to measure the inclusive yield of positive pions
from proton-beryllium collisions at 8.9 \GeVc:
\vspace{0.25ex}
\begin{equation*}
\prot + \mbox{Be} \rightarrow \piplus+ X
\end{equation*}
The absolutely normalized double-differential cross-section for this process 
can be expressed in bins of pion kinematic variables in the
laboratory frame, ($p_{\pi},\theta_{\pi}$), as 
\vspace{0.5ex}
\begin{equation}
  \frac{d^2\sigma^{\pi^{+}}}{dpd\Omega}(p_{\pi},\theta_{\pi}) = \frac{\mathrm{A}}{N_{\mathrm{A}}\cdot\rho\cdot t}\cdot\frac{1}{\Delta p\Delta\Omega}\cdot\frac{1}{N_{\mathrm{pot}}}
    \cdot N^{\pi^{+}}(p_{\pi},\theta_{\pi}) \ ,
    \label{eq:truexsec}
\end{equation}
\vspace{0.5ex}

\noindent where:
\begin{itemize}
\item  $\frac{d^2\sigma^{\pi^{+}}}{dpd\Omega}$ is the cross-section in 
\ensuremath{\mathrm{cm}^2/(\GeVc)/\mathrm{sr}}\xspace for each ($p_{\pi},\theta_{\pi}$) bin covered in the analysis
\item $\frac{\mathrm{A}}{N_{\mathrm{\tiny A}}\cdot\rho}$ is the reciprocal of the number density of
  target nuclei for beryllium ($1.2349 \cdot 10^{23}$ per \cmcube). A is the atomic mass of beryllium, $N_{\mathrm{A}}$ is Avagadro's 
  number and $\rho$ is the density of beryllium.
\item $t$ is the thickness of the beryllium target along the beam
  direction. The target has a cylindrical shape, with a measured thickness and diameter of t = ($2.046\pm 0.002$) cm and d $\simeq 3.0$ cm, respectively.
\item $\Delta p$ and $\Delta \Omega$ are the bin sizes in momentum and solid angle, respectively.\footnote{$\Delta p = p_{max}-p_{min};
\ \Delta \Omega = 2\pi(cos(\theta_{min}) - cos(\theta_{max}))$} 
\item $N_{\mathrm{pot}}$ is the number of protons on target after event
  selection cuts (see section \ref{sec:event}).
\item $N^{\pi^{+}}(p_{\pi},\theta_{\pi})$ is the yield of positive pions in bins
  of momentum and polar angle in the laboratory frame.
\end{itemize}
The true pion yield, $N^{\pi^{+}}(p_{\pi},\theta_{\pi})$, is related to the measured one,
$N^{\pi^{+^{'}}}(p_{\pi}^{'},\theta_{\pi}^{'})$,
by a set of efficiency corrections and kinematic smearing matrices. 
In addition, there is a small but non-negligible mis-identification
of particle types, predominantly between pions and protons.  
Therefore, both yields must be measured simultaneously 
in order to correct for migrations.  
Eq. \ref{eq:truexsec} can be generalized to give the inclusive
cross-section for a particle of type $\alpha$      
\vspace{0.5ex}
\begin{equation}
  \frac{d^2\sigma^{\alpha}}{dpd\Omega}(p,\theta) = \frac{\mathrm{A}}{N_{\mathrm{A}}\cdot\rho\cdot t}\cdot\frac{1}{\Delta p\Delta\Omega}\cdot\frac{1}{N_{\mathrm{pot}}}
  \cdot M^{-1}_{p\theta\alpha p^{'}\theta^{'}\alpha^{'}}\cdot N^{\alpha^{'}}(p^{'},\theta^{'}) \ ,
  \label{eq:recxsec}
\end{equation}
\vspace{0.5ex}

\noindent where reconstructed quantities are marked with a prime and 
$M^{-1}_{p\theta\alpha p^{'}\theta^{'}\alpha^{'}}$ is the
inverse of a matrix which fully describes the migrations between bins
of generated and 
reconstructed quantities, namely: laboratory frame momentum, $p$, laboratory frame
angle, $\theta$, and particle type, $\alpha$.  In practice, the matrix
$M$ can be factorized into a set of individual corrections, as will be
done here.  The reasons for doing this are threefold: 
\begin{itemize}
\item Not all efficiencies and migrations are functions of all three
  variables. Particle identification efficiencies and migrations do not
  depend on the angle, $\theta$, and the tracking efficiency and momentum 
  resolution are the same for pions and protons.
\item Using techniques described below the tracking efficiency and
  particle identification efficiency and migrations 
  can be determined from the data themselves and do not  
  rely on simulation. This is, of course, preferable wherever possible.
\item Measuring and applying the corrections separately will 
  ease the assessment of
  systematic errors as
  will be discussed in Section \ref{sec:results}. 
\end{itemize}
The form of the corrections can be separated into two basic
categories: absolute efficiencies and bin--to--bin migrations between
true and reconstructed quantities.  In particular, migrations in momentum and
in particle identification are carefully considered.  The various  
efficiency corrections can, therefore, be functions of either
reconstructed quantities or true ones, and must then be applied at the 
appropriate point in the analysis. This is important given that some
corrections, as mentioned above, are measured from the data themselves 
where one has only reconstructed quantities.  

Further, we are interested only in \emph{secondary} \piplus created in \emph{primary} 
interactions of beam protons with beryllium nuclei.  Pions created in interactions other
than p+Be at 8.9 \GeVc are a background to the measurement. \emph{Tertiary} particles are those created when
secondary particles decay or inelastically interact downstream of the target in air or detector materials and 
are not to be included in the measured cross-section.  

In the present analysis,
$M^{-1}_{p\theta\alpha p^{'}\theta^{'}\alpha^{'}}$ has been factorized  
into the following components.  Note that $\theta_x = \tan^{-1}(p_x/p_z)$ and
$\theta_y = \tan^{-1}(p_y/p_z)$ are useful variables for viewing the detector in
$x,y$-plane coordinates and are related to the standard polar angle by
$\theta = \tan^{-1}(\sqrt{\tan^{2}\theta_x + \tan^{2}\theta_y})$.
\begin{itemize}
\item
  $\varepsilon^{\mathrm{recon}}(p^{'},\theta_{x}^{'},\theta_{y}^{'})$ is
  the efficiency for the reconstruction of an `analysis track'.
  An `analysis track' is defined to include a momentum measurement as
  well as a matched time-of-flight hit needed for particle
  identification such that $\varepsilon^{\mathrm{recon}} = \varepsilon^{\mathrm{track}} \cdot \varepsilon^{\mathrm{TOFW-match}}$. 
\item $\varepsilon^{\mathrm{acc}}(\theta)$ is the correction
  for the geometric acceptance of the spectrometer and is a purely
  analytical function based on the assumption of azimuthal symmetry in
  hadron production and the fiducial cuts used in the
  analysis. See \cite{ref:alPaper} for a full description of the 
  acceptance correction and its dependence on the $\theta_y$ fiducial volume cut.
\item $M^{-1}_{pp^{'}}(\theta^{'})$ is the matrix describing the
  migration between bins of measured and generated momentum.  There is a
  unique matrix for each angular bin in the analysis since the momentum
  resolution and bias vary with angle. 
\item $M^{-1}_{\theta\theta^{'}}(p)$ is a unit matrix, implying that
  angular migrations, which are small, are being neglected. 
\item $\eta^{\mathrm{absorb}}(p,\theta_{x},\theta_{y},\alpha)$
  is the absorption plus decay rate of secondary particles before
  reaching the time-of-flight wall which is required for particle identification.
\item $(1 - \eta^{\mathrm{tert}}(p^{'},\theta_{x}^{'},\theta_{y}^{'},\alpha))$ 
  corrects for the fraction, $\eta^{\mathrm{tert}} = \frac{N^{\mathrm{rec-tert}}}{N^{\mathrm{rec}}}$,
  of total tracks passing reconstruction cuts, $N^{\mathrm{rec}}$, which are actually tertiary particles, $N^{\mathrm{rec-tert}}$.
\item $\varepsilon^{\mathrm{e-veto}}(p,\alpha)$ is the efficiency for
  particles of type $\alpha$ passing the electron veto cut used to
  remove electrons from the analysis track sample as described below.
\item $M^{-1}_{\alpha \alpha^{'}}(p)$ is the particle identification efficiency and migration matrix, assumed uniform in $\theta$.
\end{itemize}
Once again primed variables are those measured and unprimed
variables are the true quantities (i.e. after unsmearing).     
Expanding $M^{-1}_{p\theta\alpha p^{'}\theta^{'}\alpha^{'}}$ into these
individual corrections and taking care of the order in which they are
applied gives us the final equation for calculating the absolute
cross-section from the measured yields.
\vspace{0.5ex}
\begin{equation}
  \begin{split}
    \frac{d^2\sigma^{\alpha}}{dpd\Omega}(p,\theta) = & \frac{\mathrm{A}}{N_{\mathrm{A}}\cdot\rho\cdot t}\cdot\frac{1}{\Delta p\Delta\Omega}\cdot\frac{1}{N_{\mathrm{pot}}} \\
    & \times M^{-1}_{\alpha \alpha^{'}}(p)
    \cdot \frac{1}{\varepsilon^{\mathrm{e-veto}}(p,\alpha)} \\
    & \cdot \frac{1}{1 - \eta^{\mathrm{absorb}}(p,\theta_{x},\theta_{y},\alpha)} \\
    & \cdot M^{-1}_{pp^{'}}(\theta^{'})
    \cdot (1-\eta^{\mathrm{tert}}(p^{'},\theta_{x}^{'},\theta_{y}^{'},\alpha)) \\
    & \cdot \frac{1}{\varepsilon^{\mathrm{acc}}(\theta^{'})}
    \cdot \frac{1}{\varepsilon^{\mathrm{recon}}(p^{'},\theta_{x}^{'},\theta_{y}^{'})} \\
    & \cdot N^{\alpha^{'}}(p^{'},\theta^{'}) \ .
    \label{eq:finalxsec}
  \end{split}
\end{equation}
\vspace{0.5ex}

There are two additional aspects of the analysis methods which are
worth mentioning.  First, particle distributions are built by
multiplying a set of  
correction weights for each reconstructed track and weighting
events before they are added to the total yields.  In this way a
single reconstructed 
track is 'spread' over multiple true momentum bins according to the
elements of $M^{-1}_{pp^{'}}(\theta^{'})$, and the population in each
true bin is comprised of tracks from all reconstructed momentum bins.
This approach avoids the difficulties associated with inverting a large smearing matrix due to 
potential singularity of the matrix as well as potential pathologies in the inverted matrix caused by a loss of
information at the kinematic boundaries of the matrix itself.  The drawback to this method is that one has some 
sensitivity to the underlying spectrum in the Monte Carlo used to generate the matrix (see Sec. \ref{sec:momentum}). 
    
Second, there is a background associated 
with beam protons interacting in materials other than the nuclear target (parts of the detector, air, etc.).  
These events can be subtracted by 
using data collected without the nuclear target in place.
We refer to this as the 'empty target subtraction':  
\vspace{0.5ex}
\begin{equation*}
  N^{\alpha^{'}}(p^{'},\theta^{'}) \rightarrow [N^{\alpha^{'}}_{\mathrm{target}}(p^{'},\theta^{'}) - N^{\alpha^{'}}_{\mathrm{empty}}(p^{'},\theta^{'})] \ .
\end{equation*}
\vspace{0.5ex}

The final form of the cross-section calculation is then given by making the above substitution into Eq. \ref{eq:finalxsec}.

\section{Track selection and reconstruction efficiency corrections}\label{sec:analysistracks}


\subsection{Event selection}\label{sec:event}

\begin{figure*}[ht]
  \begin{center}
    \includegraphics[width=17cm,height=5cm]{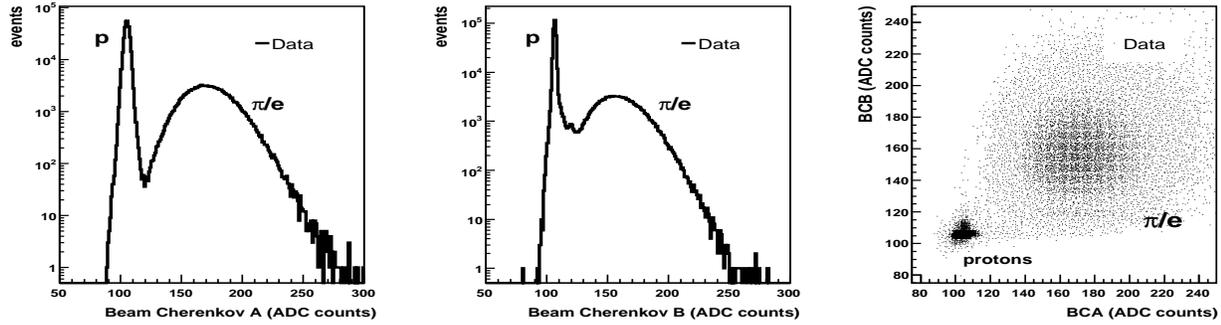}
    \caption{\label{fig:beamckov}  
      Beam Cherenkov pulse height distributions. BCA in the left panel, BCB in the middle, and BCB \emph{vs.} BCA in the right panel. The electron
      and pion tagging efficiency is found to be close to 100\%; the peaks are separated by $\approx 3\sigma$ in both detectors. By requiring a 
      value compatible with a pedestal in both Cherenkov detectors the beam protons are clearly separable from pions and electrons 
      as seen in the right panel.}
  \end{center}
\end{figure*}

\begin{figure}[ht!]
  \begin{center}
    \includegraphics[width=9cm,clip=true,trim=0.4cm 0cm 0cm 0cm]{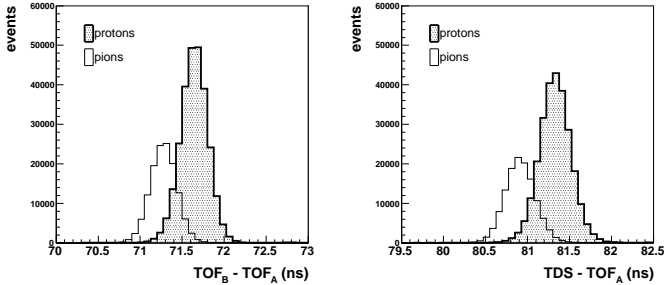}
    \caption{\label{fig:beamtof}
      Beam time-of-flight distributions. The time difference between
      TOFA and TOFB is shown in the left panel.  The right panel is
      the  
      time difference between TOFA and the TDS.  The shaded
      distributions are for particles identified as protons by the
      Cherenkov detectors 
      as described in the text. The open histograms are all other beam
      tracks: pions, electrons and muons from pion decays.} 
  \end{center}
\end{figure}

Protons are identified in the T9 beam at 8.9~\GeVc exactly as in the
12.9~\GeVc data set and as described in reference \cite{ref:alPaper}. 
Two threshold Cherenkov detectors (BCA and BCB) placed in the beam line are used to select protons by requiring
a value consistent with the pedestal in both detectors.  
The beam Cherenkov pulse height distributions for the 8.9 \GeVc beam are shown in 
Fig. \ref{fig:beamckov}.  Protons were selected by requiring a pulse height less than 120 counts in both detectors, and
Fig. \ref{fig:beamtof} shows the time-of-flight distributions of those beam tracks 
identified as protons and pions by the Cherenkov selection. The beam time-of-flight system is made of 
two identical scintillator hodoscopes, TOFA and TOFB, recuperated from the previous NA52 experiment and a 
small target-defining trigger counter (TDS).  TOFA-TOFB and TOFA-TDS measure time differences over a distance of
21.4 m and 24.3 m, respectively.  We see in Fig. \ref{fig:beamtof} that the two time peaks are consistent with the proton
and pion hypotheses at 8.9 \GeVc.  

Only events with a single reconstructed beam track in the four beam 
multi-wire proportional chambers (MWPCs) and no signal in the
beam halo counters are accepted.  This MWPC track is used to determine the impact position and angle 
of the beam particle on the target.  A time measurement in one of three 
beam timing detectors consistent with a beam particle is also required
for determining the arrival time of the proton at the target, $t_0$.
This $t_0$ is necessary for calculating the time-of-flight
of secondary particles.

The full set of criteria for selecting beam protons for this analysis is as follows:
\begin{itemize}
  \item ADC count less than
  120 in both beam Cherenkov A and beam Cherenkov B 
  \item time measurement(s) in TOFA, TOFB and/or TDS which are needed for
  calculating the arrival time  of the beam proton at the target, $t_0$
  \item extrapolated position at the target within a 10~\mm radius of the center of the target
  \item extrapolated angle at the target less than 5 mrad
  \item no signal in the beam halo counters
\end{itemize}

Prior to the above cuts, for data taken with a nuclear target, a downstream trigger in the forward trigger plane (FTP) 
was required to record the event.\footnote{empty target data sets are recored with an unbiased trigger setting since
these samples are used to calibrate the experimental apparatus and not just in
the empty target subtraction for cross-section measurements.}  
The FTP is a double plane of scintillation counters
covering the full aperture of the spectrometer magnet except a 60 mm central hole for allowing
non-interacting beam particles to pass.  The efficiency of the FTP is 
measured to be $>$99.8\%. 

Using the FTP as an interaction trigger necessitates an additional set of unbiased,
pre-scaled triggers for absolute normalization of the  
cross-section.  Beam protons in the pre-scale trigger sample (1/64 of the total trigger rate for the 8.9 \GeVc Be
data set) are subject to exactly the same selection 
criteria as FTP trigger events allowing the
efficiencies of the selections to cancel and adding no additional
systematic uncertainty to the absolute normalization of the result.  These unbiased events are 
used to determine the $N_{\mathrm{pot}}$ used in the cross-section formula and listed in Table \ref{tb:be5events}. 
The number of protons-on-target is known to better than 1\%.  

Applying these criteria we are left with the event totals summarized in Table \ref{tb:be5events}.

\subsection{Secondary track selection}\label{sec:track}

The following criteria have been applied to select tracks in the forward spectrometer for the  
accepted events:
\begin{itemize}
  \item a successful momentum reconstruction using downstream track
    segments in NDC modules 2, 3, 4 or 5 and the position  
    of the beam particle at the target as an upstream constraint
    (here, \emph{up}stream and \emph{down}stream are relative to the
    spectrometer magnet); 
  \item a reconstructed vertex radius ({\em i.e.} the distance of the
  reconstructed track from the $z$-axis in a plane perpendicular to
  this axis at $z=0$) $r \ \leq$ 200 \mm;
  \item number of hits in the road around the track in NDC1 $\geq$ 4 
    and average $\chi^2$ for these hits with respect to the track in
    NDC1 $\leq$ 30
    (this is applied to reduce non-target interaction backgrounds); 
  \item number of hits in the road around the track in NDC2 $\geq$ 6
  (this is applied to reduce non-target interaction backgrounds); 
  \item a matched TOFW hit passing the quality cuts described in
  Sec. \ref{sec:tofhitselect}; 
  \item reconstructed angles are within the fiducial volume to be used
  for this analysis, $-$210~mrad $\leq \theta_{x} \leq$ 0~mrad and  
    -80~mrad $\leq \theta_{y} \leq$ 80~mrad.
\end{itemize}
These cuts are identical to those used in the analysis of the p-Al data except for the 
reconstructed vertex radius $\leq200$ \mm cut. It was found that due to a feature of the algorithm 
this additional requirement improved the momentum resolution considerably at reconstructed momenta below $\approx$1.5 \GeVc.

Applying these cuts to reconstructed tracks in accepted events we are
left with 95,897 total good tracks in the beryllium thin target data
set as listed in Table \ref{tb:be5events}. 

\begin{table*} 
\label{tb:be5events}
\begin{center}
\begin{tabular}{|l c c|} \hline
\bf{Data Set} & \bf{Be 5\% 8.9 \GeVc} & \bf{8.9 \GeVc Empty Target}\\ \hline
    protons on target                & 13,074,880 & 1,990,400 \\
    total events processed           & 4,682,911 & 413,095\\ 
    events with accepted beam proton & 2,277,657 & 200,310 \\ 
    beam proton events with FTP trigger & 1,518,683 & 91,690 \\
    total good tracks in fiducial volume & 95,897  & 3,110 \\ \hline
\end{tabular}
\vspace{3ex}
\caption{Total number of events in the 8.9~\GeVc beryllium
  5\%~$\lambda_{\mathrm{I}}$ target 
  and empty target data sets, and the number of protons on target as
  calculated from the prescaled trigger count.} 
\end{center}
\end{table*}


\begin{figure*}[!htb]
  \begin{center}
    \includegraphics[width=11.5cm]{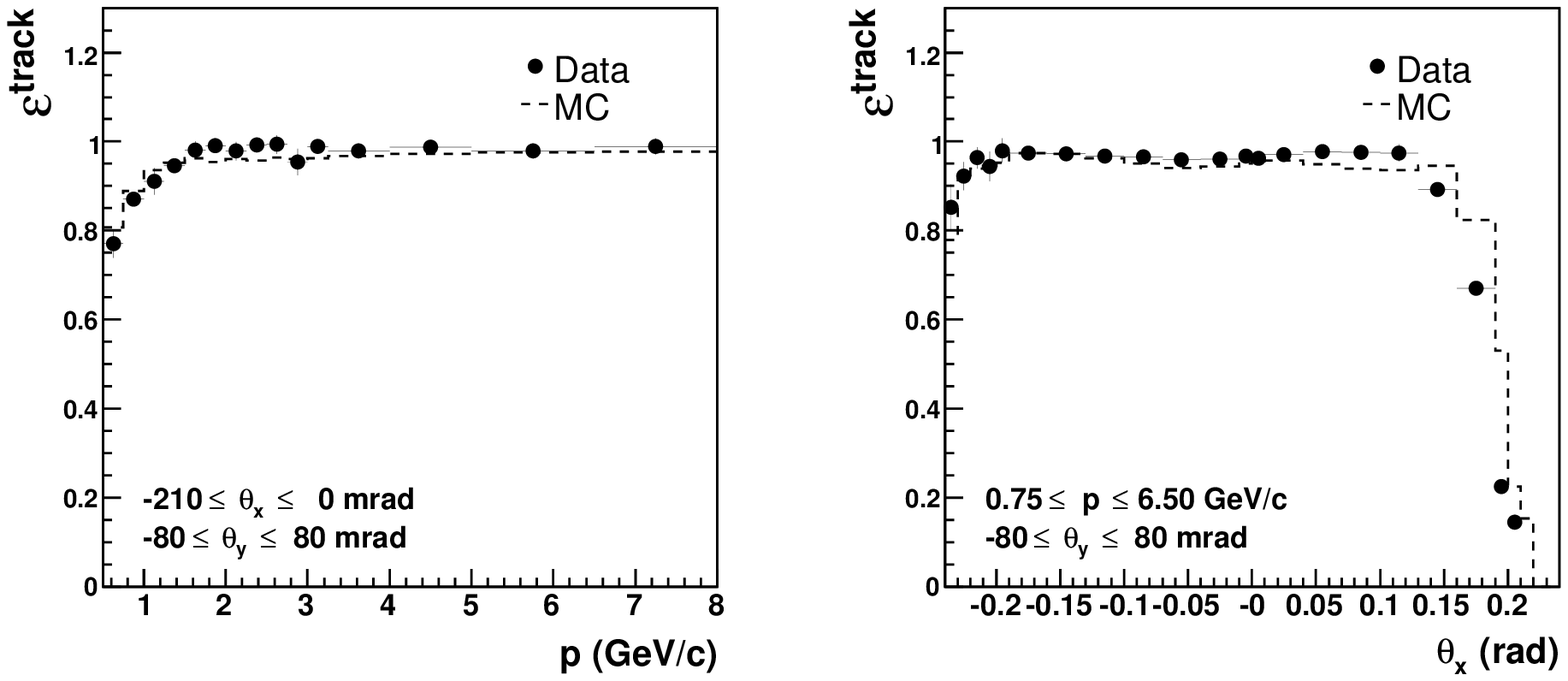}
    \includegraphics[width=5.75cm]{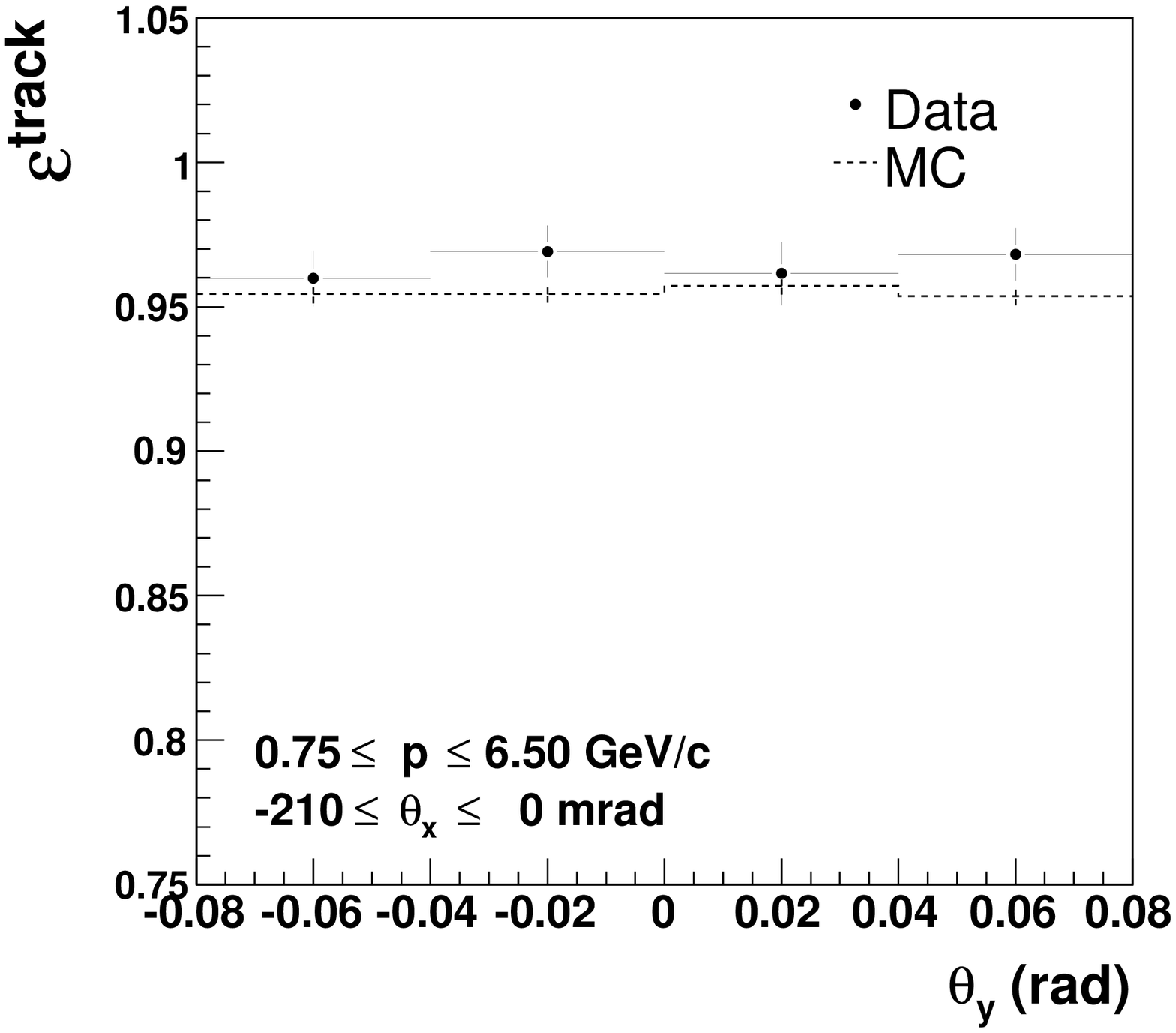}
    \caption{\label{fig:trackEff}  Tracking efficiency for positive particles traversing 
      the detector using the target as upstream track constraint as a function of particle
      momentum (upper left), production angle in the horizontal plane, $\theta_x$ (upper right), and
      production angle in the vertical plane, $\theta_y$ (lower). 
      The $\theta_{y}$ plane is orthogonal to the spectrometer bending plane and not sensitive to the momentum
      dependent acceptance, and the bottom panel shows the purest measure of the average 
      track reconstruction efficiency within the fiducial volume to be 96\%--97\%.}
  \end{center}
\end{figure*}

\subsection{Track reconstruction efficiency}\label{sec:reconEff}

The track reconstruction efficiency has been measured from the data exactly as described in \cite{ref:alPaper}.
The efficiency is shown in Fig. \ref{fig:trackEff}.  The effects of the two changes in track reconstruction are
evident in the efficieny curves.  First, the efficiency is now flat and $\approx 97$\% above 
2 \GeVc due to the improvement in the $\chi^2$ minimization done as part of the tracking algorithm. 
Second, the loss of efficiency at momenta below 1.5 \GeVc 
is due to the reconstructed vertex radius $\leq200$ \mm cut discussed above.  
But, as before, the tracking efficiency can be
measured from the data themselves, so the systematic error on the correction comes only from the statistical uncertainty in
the sample used to calculate the correction.  As in the p-Al publication, the 12.9 \GeVc aluminum data and the 8.9 \GeVc
beryllium data have been combined to minimize this uncertainty.
The drop in efficiency at large, positive values of $\theta_x$ is due to geometric acceptance as low momentum tracks are
bent out of the spectrometer missing the downstream chambers. The present analysis is performed using 
tracks in the range $-0.210 \ \rad \leq \theta_x \leq 0 \ \rad$ where the acceptance is flat in momentum.

\subsection{Absorption, decay and tertiary track corrections}\label{sec:absorption}

\begin{figure*}
  \begin{center}
    \includegraphics[width=17cm,height=5.5cm]{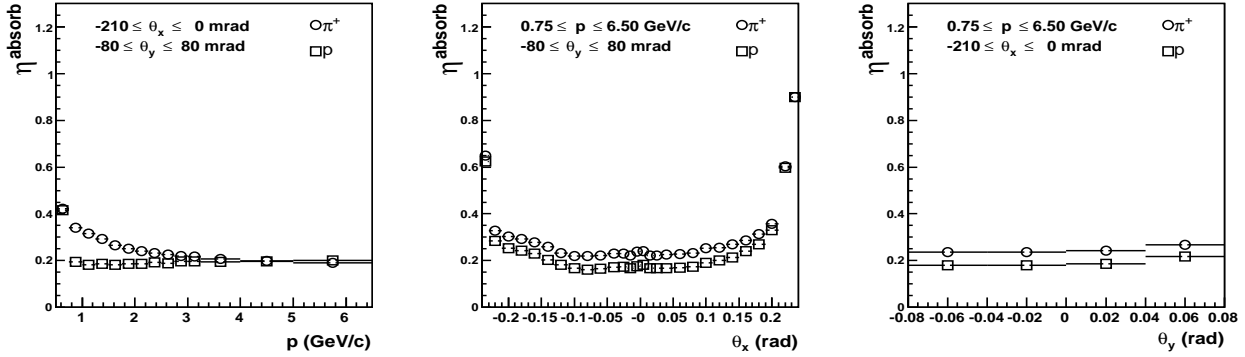}
    \caption{\label{fig:absorption}
      Absorption corrections for pions and protons according to Monte Carlo simulation as a function of particle
      momentum (left), production angle in the horizontal plane, $\theta_x$ (center) and production angle in 
      the vertical plane, $\theta_y$ (right). 
    }
  \end{center}
\end{figure*}

The correction for absorption and decay refers to secondary particles created in the nuclear target that 
never make it to the time-of-flight wall for detection and possible identification.  Figure \ref{fig:harpDet} shows the location
of the time-of-flight scintillator wall just beyond the back plane of drift chamber modules NDC3, NDC4 and NDC5.  
We use the Monte Carlo simulation 
to determine the size of the correction and the result is shown in Fig. \ref{fig:absorption}. 
Note this is an upward adjustment to the raw yield measured and is 
implemented as $1/(1-\eta^{absorb}(p,\theta_{x},\theta_{y},\alpha))$ in Eq. \ref{eq:finalxsec}.  The absorption correction
(which includes pion decays) is a function of $\theta_{x}$ and $\theta_{y}$ because it depends on the amount and type
of physical material a particle passes through, thus the geometry of the detector.  It is a function of $\alpha$ because
of the different interaction cross-sections and possible decay rates of hadrons.  
This correction is separated from the tertiary
correction discussed below because it does not depend on event multiplicity, kinematics or other details of the
 hadron production model used in the simulation, but only the total interaction cross-sections 
which are significantly more certain.  In fact, the relevant cross-sections are typically known to $\approx10$\% and we 
assume this uncertainty on the absorption correction just as in the p-Al publication.  Because the correction is of order 30--40\% for pions, 
the average systematic error contribution to the cross-section turns out to be 3.6\%.

\begin{figure*}
  \begin{center}
    \includegraphics[width=17cm,height=5.5cm]{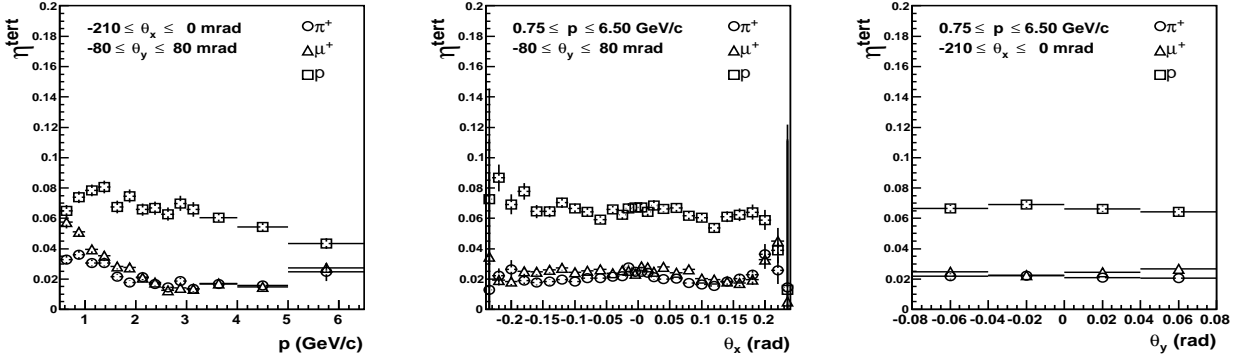}
    \caption{\label{fig:tertiary}
      Tertiary particle rates for pions, muons (which get identified as pions) and protons according to Monte Carlo simulation 
      as a function of particle
      momentum (left), production angle in the horizontal plane, $\theta_x$ (center) and production angle in 
      the vertical plane, $\theta_y$ (right).}
  \end{center}
\end{figure*}

The tertiary correction refers to the subtraction of reconstructed tracks which are actually reconstructions
of tertiary particles, \emph{i.e.} particles produced in inelastic interactions or decays of true secondary particles
and not in primary interactions of 8.9 \GeVc protons with beryllium nuclei (See section \ref{sec:xsec}).   The tertiary subtraction includes muons
created in decays which are falsely identified as pions nearly 100\% of the time due to their high $\beta$. 
The correction is significantly smaller than the absorption correction (compare Figs. \ref{fig:absorption} and \ref{fig:tertiary}), 
but is less certain, so the contribution to the systematic error is non-negligible.
This tertiary subtraction is also generated using the Monte Carlo simulation but \emph{is} dependent on the details of the
hadron production model used in the simulation. Most of the material where tertiary 
particles might be produced in the detector is carbon, so it is the simulation of inelastic interactions of low-energy protons and 
pions in carbon that become important in generating this correction.  Previously this correction was assumed to be 100\% 
uncertain, but comparisons of low momentum HARP p+C, $\pi^{+}$+C and $\pi^{-}$+C data to the hadronic models used in the simulation
have verified these models to $\approx 50$\% and allowed us to lower the systematic error on this correction.  Fig. \ref{fig:tertiary}
shows the average size of the correction to the \piplus yield to be about 4\% (2\% $\pi^{+}$ + 2\% $\mu^{+}$), 
so the average systematic error on the cross-section coming from the subtraction of tertiaries ends up being 1.8\%.

\section{Momentum resolution and scale corrections}\label{sec:momentum}

Two important sources of systematic uncertainty are the momentum resolution and the
absolute momentum scale of the reconstruction.  Each, in general, vary with the value
of the true momentum and angle of the track.  The Monte Carlo will be used 
to generate corrections to the measured momenta so a validation of the simulation becomes necessary.     
Three techniques have been developed to compare the resolutions of reconstructed quantities 
in data and Monte Carlo of which only one was available at the time of the p-Al paper.  
The challenge is to isolate a set of tracks in the data sample with a known
momentum.  The three methods are described fully below but, briefly, they are based on empty target data sets, 
samples of elastic scattering events and using the excellent resolution of the time-of-flight system 
to determine the momentum.  Results from these studies have been 
used to tune the detector simulation and to estimate
a systematic uncertainty due to momentum resolution and scale.

A multiplicative momentum scale correction is applied to all reconstructed tracks in the data to 
remove a $\theta_x$, $\theta_y$ dependence seen in calibration samples.
After this correction we see no significant momentum mis-calibration 
beyond the 2\% absolute momentum scale uncertainty estimated using the elastic scattering technique
(see Sec. \ref{sec:fw-elastic}). The minimum ratio of momentum bin width over momentum bin central value 
is 8\%, four times this value.
An average uncertainty on the \piplus cross-section of 3.6\% is estimated from this effect (the
effect is dependent on the steepness of the pion and proton spectra, the size of the momentum bias and the momentum binning).

The resolution of the momentum measurement leads to smearing in the reconstructed pion spectrum.   
Momentum `unsmearing' is performed in the analysis by using the Monte
Carlo to generate a momentum migration matrix,
$M^{-1}_{pp^{'}}(\theta^{'})$ which describes how a reconstructed
momentum value is distributed across bins of true momentum.  The accuracy
of this migration matrix, which is generated from Monte Carlo samples,
depends upon two factors -- the similarity of the underlying
hadron distributions to those in the data and the agreement between
the resolutions of reconstructed quantities in the data and Monte
Carlo relative to the true momenta.  The latter,
the agreement between the resolutions of reconstructed quantities in
data and  Monte Carlo, is the topic of the next three subsections.  The techniques developed have been 
used to tune parameters of the simulation to achieve good agreement.  The first effect, 
the similarity of the hadron spectra to the data, therefore, will dominate the systematic error for this correction. 
We estimate a momentum resolution uncertainty by performing the cross-section analysis using
migration matrices generated from Monte Carlo samples which use different hadronic models.  
The average uncertainty on the \piplus cross-section estimated from this effect is 3.4\%.

\subsection{Momentum calibration using empty target data sets}
\label{sec:fw-empty}
\begin{figure}
  \begin{center}
    \includegraphics[width=8cm]{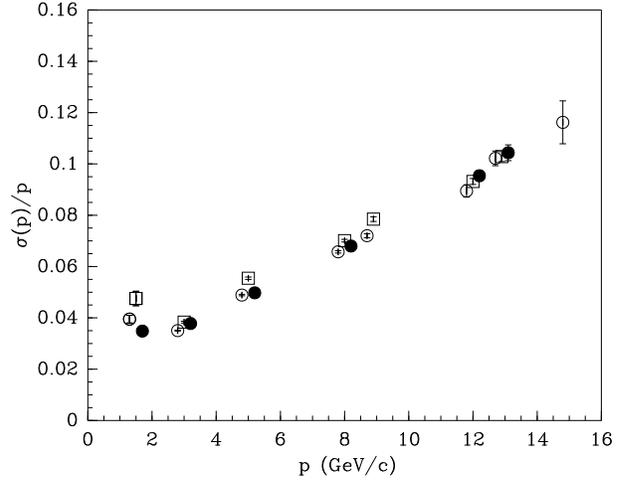}
    \caption{\label{fig:momResEmpty}
      Momentum resolution $\sigma(p)/p$ (Gaussian fit) as a function of momentum
      (in \GeVc) for the drift chambers: the data were taken using several well-defined discrete
      beam momenta and no target for pions (open circles) and protons (open squares).
      Also shown (filled circles) is the corresponding resolution found using the Monte Carlo simulation.
    }
  \end{center}
\end{figure}

The first method for calibrating the momentum reconstruction 
uses empty target data sets where the
incoming beam momentum value acts as the known momentum.
Fig. \ref{fig:momResEmpty} shows the result of such a study using
empty target data and Monte Carlo samples for beam momenta of 1.5,
3.0, 5.0, 8.0, 8.9, 12.0, 12.9 and 15.0 \GeVc.  The shape of the
$\sigma_p/p$ curve versus momentum is as expected, and the
agreement between data and Monte Carlo is excellent.
For the previous analysis \cite{ref:alPaper} only these `test-beam data'
were available, sampling the spectrometer response on the $z$-axis only.
Now, additional methods have been used.
A method using samples of elastic scattering events extends the range to larger angles while a method
employing the time-of-flight detector covers the region of low
momenta.  Thus the full range of relevant angles and momenta is covered.

\subsection{Momentum calibration using elastic scattering events}
\label{sec:fw-elastic}
The elastic scattering process provides one track in the forward
direction with a momentum close to the beam momentum and a soft large
angle proton.
In p--p and $\pi$--p scattering the kinematics is fully determined by the
measurement of 
the direction of one of the outgoing particles. 
The precision of the measurement of the angle of the forward scattered
particle is sufficient to predict the momentum of that particle without
significant error.
Elastic events can be readily selected by imposing combined criteria
in the large angle and forward spectrometer.
The main selection is the requirement of one and only one track in the
TPC, identified as a proton, and exactly one track in the forward
direction. 
Further constraints have been used, such as a match of the kinematics
of these tracks and the selection of events with exactly the expected
number of hits in the trigger counters and the barrel RPC detectors
around the TPC.
In these selections no constraints on the momentum measurement of the
forward track has been set.
The purity of the selection of the elastic scattering events used in this analysis 
can be estimated by studying the $p$--$\theta$ distribution of the proton recoil 
tracks measured in the TPC.  The two quantities are fully correlated for elastic scattering events.  
From the small number of events outside the expected region, one can estimate the background 
in the sample not to exceed 1\%.  The systematic error introduced by this 
background is negligible compared to the 2\% estimated overall momentum calibration error. 

\begin{figure}[tbp]
  \begin{center}
    \includegraphics[width=4.3cm,height=5cm,clip=true,trim=0cm 0cm -1.2cm 0.7cm]{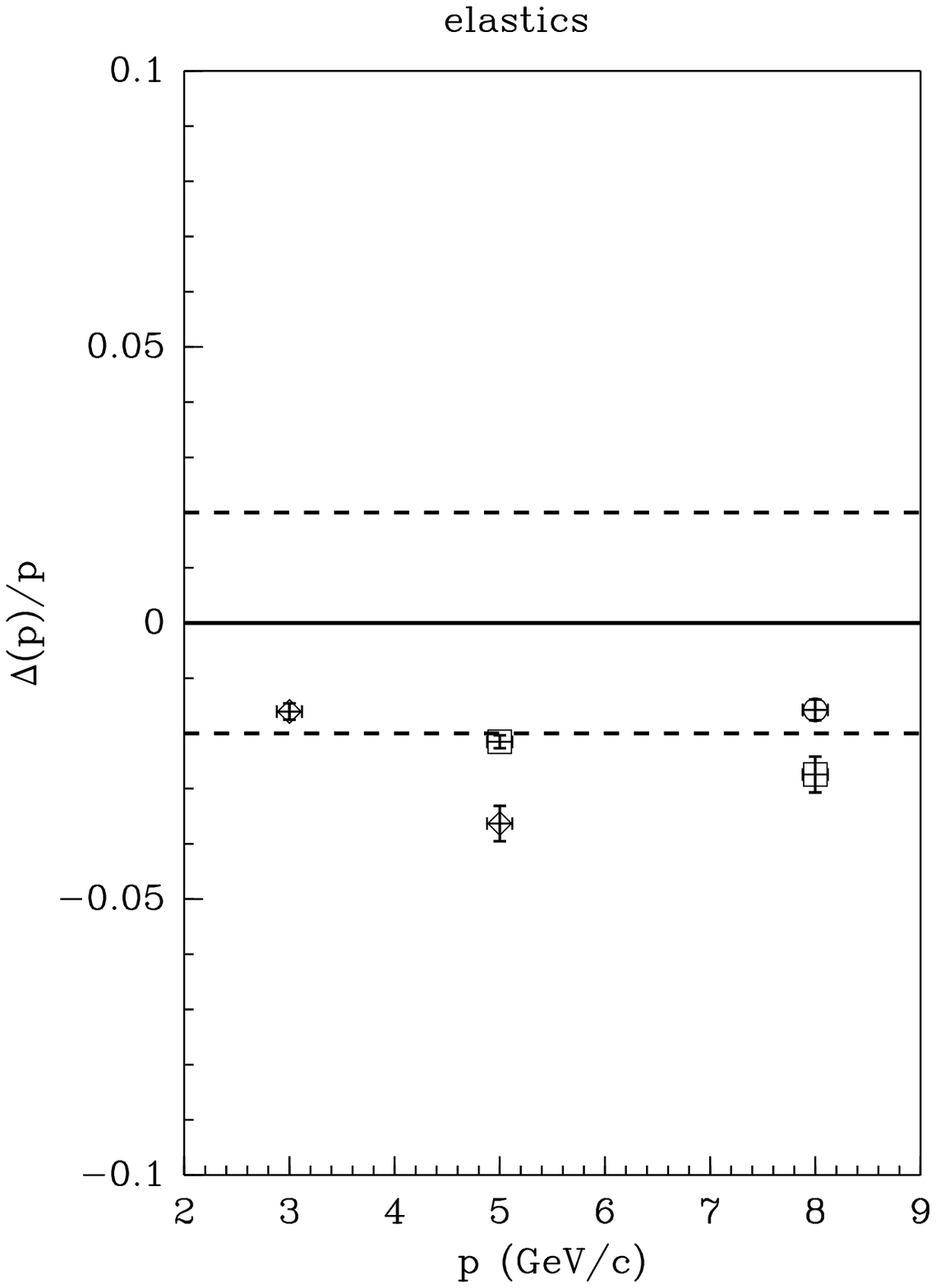}
    \includegraphics[width=4.3cm,height=5cm,clip=true,trim=-0.3cm 0cm -1.2cm 0.5cm]{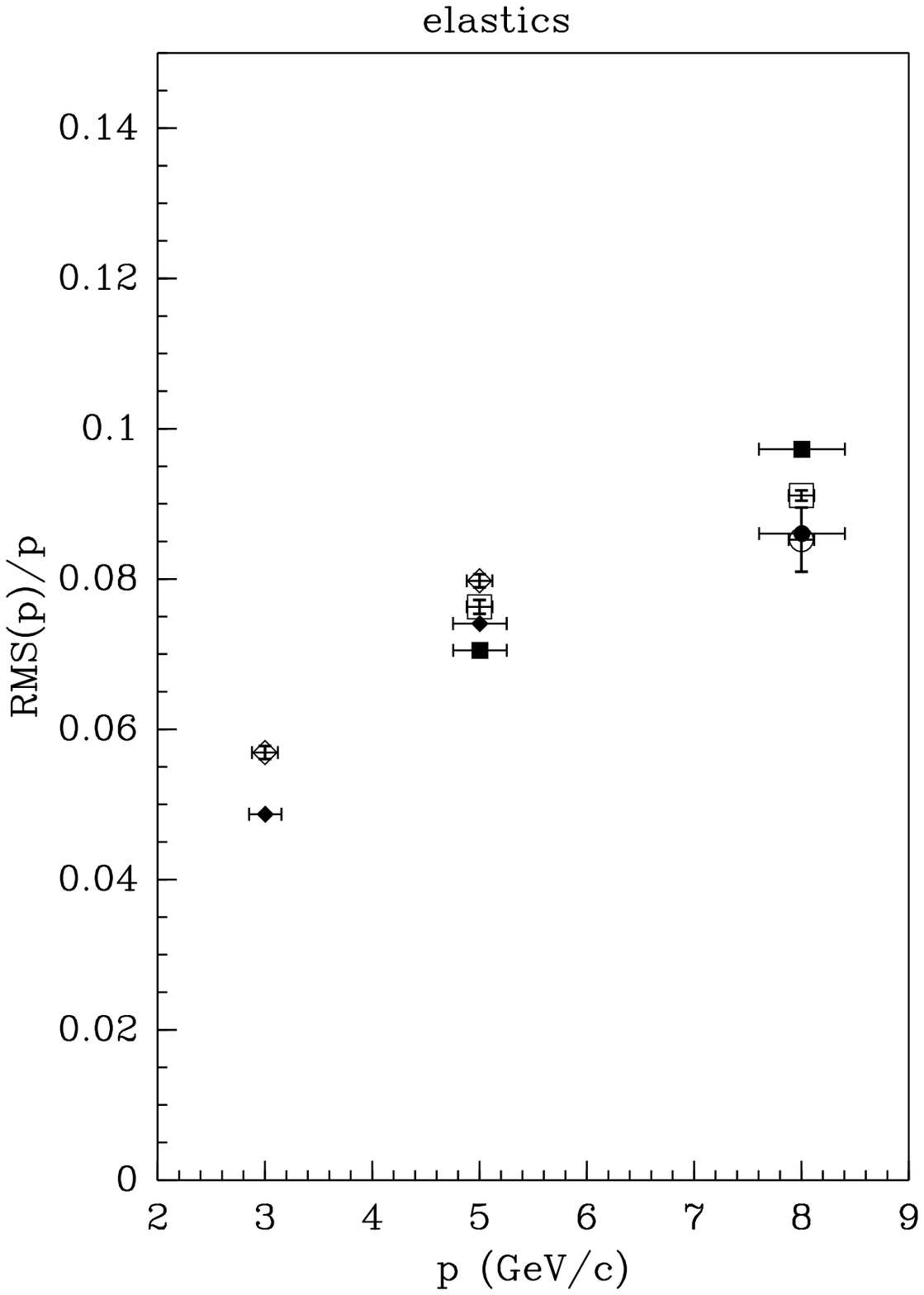}
    \caption{\label{fig:momResElas}
      Momentum scale $\Delta(p)/p$ and resolution $\sigma(p)/p$ (Gaussian fit) as a function
      of momentum in three angular regions (circles: 30~\mrad--60~\mrad; boxes: 60~\mrad--100~\mrad; diamonds 100~\mrad--150~\mrad). 
      The data were taken using 3~\GeVc,  5~\GeVc and 8~\GeVc 
      beams impinging on a hydrogen target.  The top panel shows no significant momentum mis-calibration 
      beyond the 2\% absolute momentum scale uncertainty estimated using the elastic scattering technique.
      The bottom plot shows the resolutions from data (open symbols) 
      compared with the corresponding Monte Carlo simulations (filled symbols).
      Incoming pion and proton data are combined.
    }
  \end{center}
\end{figure}

Figure \ref{fig:momResElas} shows the 
results of an analysis of elastic events using 3~\GeVc,  5~\GeVc,
and 8~\GeVc beams impinging on a hydrogen target. 
Incoming pion and proton data are combined.
The left panel reveals a momentum offset in the data of about 2\% at all momenta 
and in three angular regions, 30~mrad--60~mrad, 60~mrad--100~mrad and 100~mrad--150~mrad.  This 2\% will contribute 
a systematic uncertainty to the final cross-section and will be discussed in Sec. \ref{sec:errors}.
The right panel shows the RMS of the momentum measurement as a function of momentum
in the same two angular regions.  The resolution measured from data is compared 
to that in the Monte Carlo and, as with the empty target samples, the agreement is excellent.

The momentum scale and resolution measured with the elastic events
extends the calibration toward larger angles than probed with the
empty target data alone and make it possible to characterize the
spectrometer over a larger range of its aperture.
However, the lowest momentum at which these data are available is
3~\GeVc, such that another method had to be developed to study the
performance of the spectrometer for lower momentum particles.

\subsection{Momentum calibration using time-of-flight}

The TOFW system \cite{ref:tofPaper} can be used to provide a momentum calibration in a
momentum range where the dependence of the time-of-flight on the
momentum for pions is 
much smaller than for protons and at the same time the resolution of
the TOFW is better than the prediction of $\beta$ for protons based on
the momentum measurement.
For protons the sensitivity to the momentum resolution is larger
than all other dependencies in the beta resolution in the region 
$0.22~\GeVc < p < 2.0~\GeVc$. 
The TOFW$\beta$  resolution is typically $\sigma(\beta)/\beta=0.005$.
The precision 
$\sigma(\beta_{\mbox{pred}})/\beta$ in the
calculation of $\beta$ for protons ranges from 0.05 at $0.22~\GeVc$ to
0.005 at $2~\GeVc$ due to the momentum resolution which is of the
order of \\ 100~\MeV/c in this range.
Therefore the width of the $\beta$ peak for a sample of protons
selected in a small range in {\it measured momentum} shows a large
sensitivity to the momentum resolution.

To exploit this feature for the determination of the momentum
resolution at small momenta the time-response of the TOFW has to be
measured for pions and protons separately.
First a very clean sample of pions is selected.
Particles of negative charge are selected for this purpose to provide
negative pions.
In principle, this sample can be contaminated by electrons and
negative kaons.
Antiprotons are expected to be negligible.
At a momentum below the Cherenkov threshold for pions, electrons are
rejected by retaining only particles without a signal in the Cherenkov 
detector. 
The remaining sample contains mainly \pim with a small background of
\kam. 
This kaon background is visible in the TOF spectrum and has been taken
into account.
\begin{figure}[!h]
  \begin{center}
    \includegraphics[width=4.2cm,height=4cm,trim=0cm -.75cm 0cm 0cm]{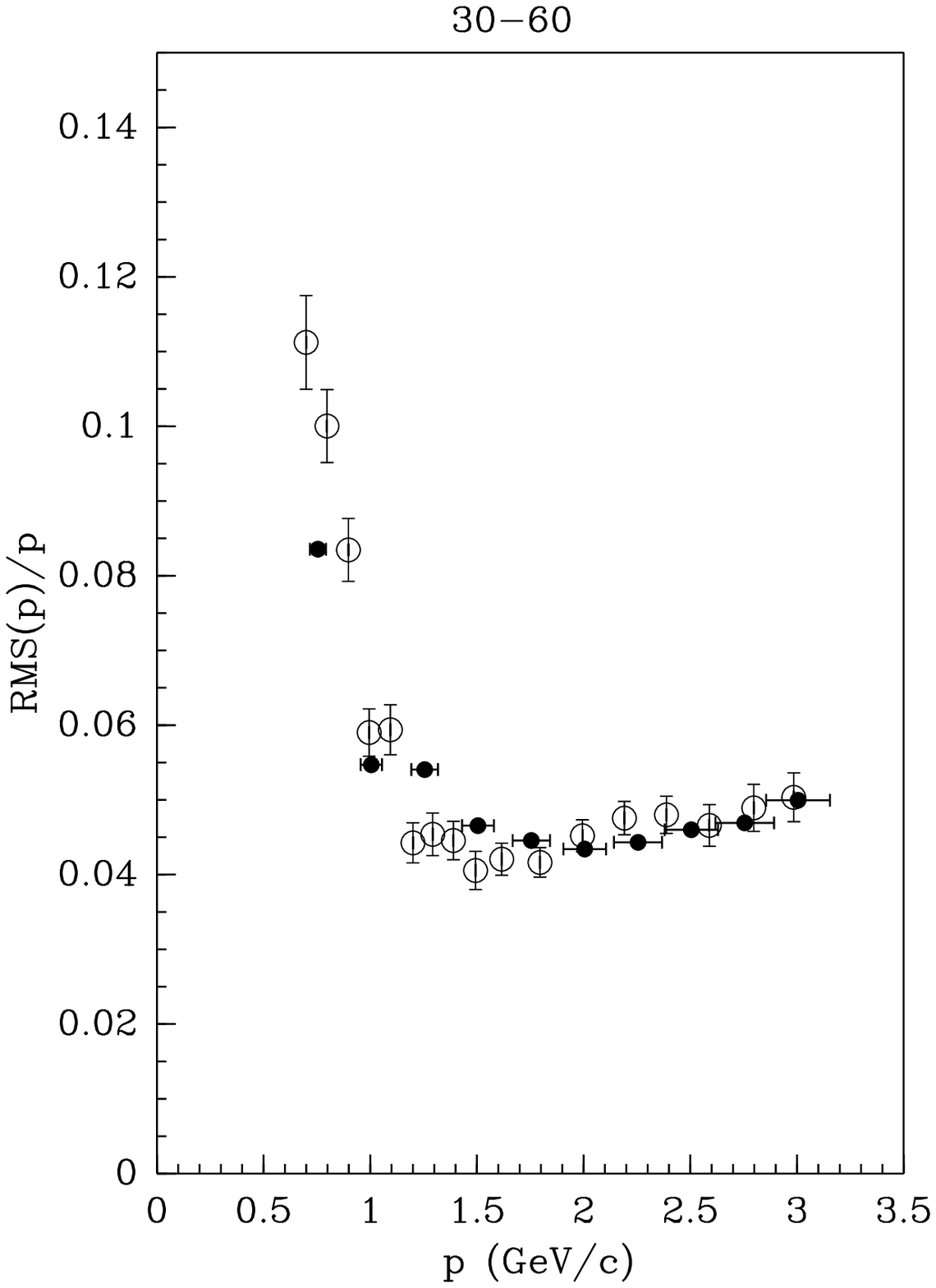}
    \includegraphics[width=4.2cm,height=4cm,trim=0cm -.75cm 0cm 0cm]{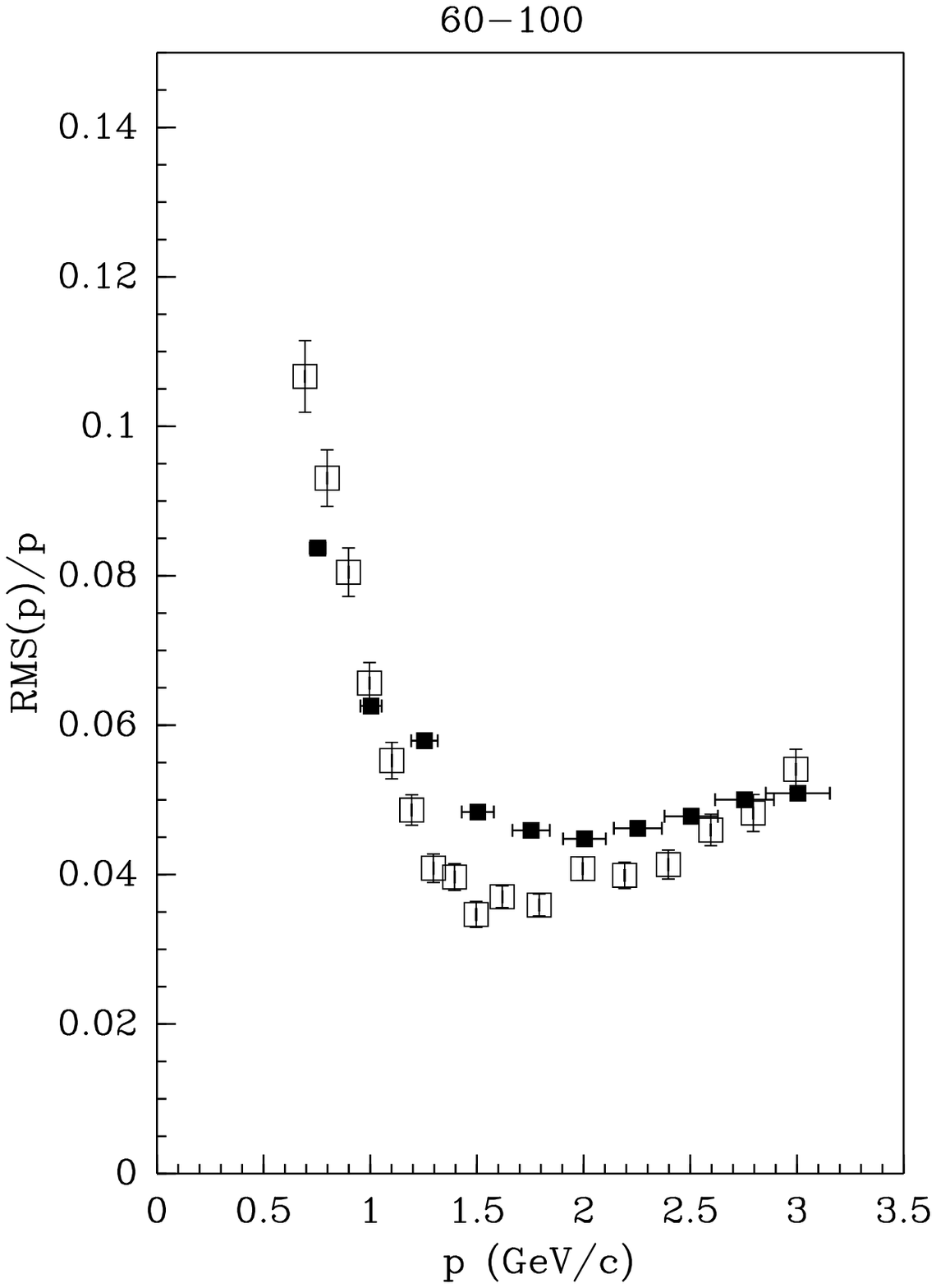}
    \includegraphics[width=4.2cm,height=4cm,trim=0cm 0cm 0cm -.75cm]{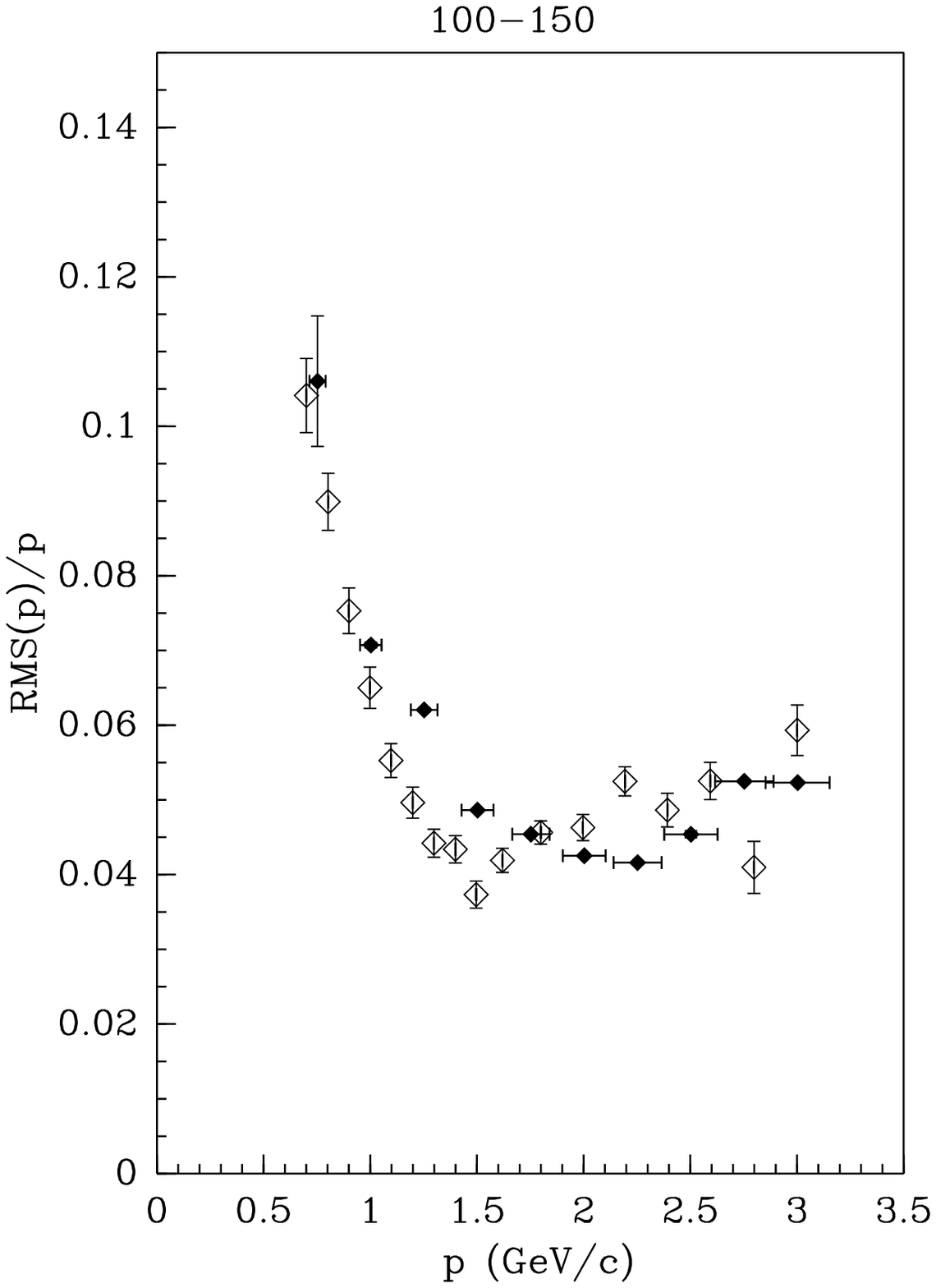}
    \includegraphics[width=4.2cm,height=4cm,trim=0cm 0cm 0cm -.75cm]{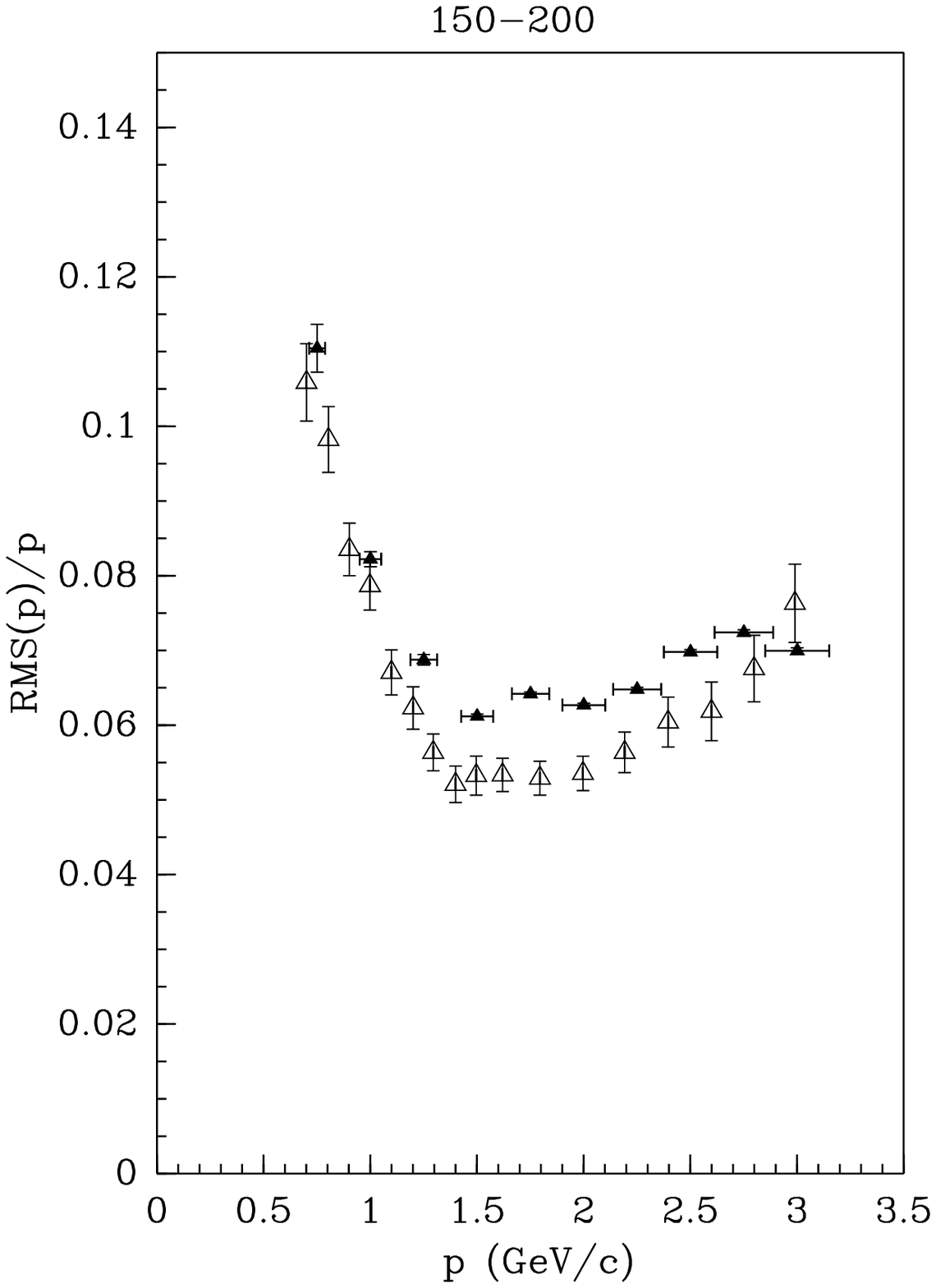}
  \end{center}
  \caption[]{
    RMS momentum resolution as a function of momentum in four angular bins
    (circles: 30 \mrad--60 \mrad; boxes: 60 \mrad--100 \mrad; diamonds:
    100 \mrad--150 \mrad; triangles: 150 \mrad--200 \mrad).  Data are shown with open symbols; 
    Monte Carlo with filled symbols.
  }
  \label{fig:betamethod}
\end{figure}

At a momentum above the pion Cherenkov threshold, the \pim sample is
selected by requiring a signal in the Cherenkov.
Although the electron background is expected to be small at these
momenta, a selection with the calorimeter is
used to reject them.  
The same selection is also used at momenta below Cherenkov threshold
in addition to the rejection with the Cherenkov detector.
Thus a sufficiently clean sample of negative pions can be obtained in
the whole momentum range.

From the sample of positive particles, positrons are removed by the
selection using the electron identifier.
Below Cherenkov threshold only particles without signal
in the Cherenkov are used. 
This selection retains protons, kaons and pions below pion Cherenkov
threshold and only protons and kaons above the threshold.
The precision of the TOFW is sufficient to provide a good separation of
pions and protons below pion \\ Cherenkov threshold.
Thus, in the whole range relevant for the experiment a clean sample
of protons can be obtained, albeit with a small contamination of kaons.
As mentioned above, the measurement with the negative pions
characterizes the TOFW response, both its absolute time and its time
resolution. 
The negative pions also provide a perfect prediction for the behavior
of the TOFW measurement for the positive pions.

The measurement of the properties of the momentum determination in the
spectrometer is then obtained by selecting small regions (bins) of
measured momentum and fitting the $\beta$ spectrum of
protons with a function which takes into account the width of the
momentum bin,
the calibrated $\beta$ resolution and as a free parameter the momentum
resolution. 
The results for data and Monte Carlo and in four angular regions
are shown in Fig.~\ref{fig:betamethod}.
The resolution measured should be interpreted as an RMS of the
momentum resolution and is larger than the $\sigma$ of the Gaussians
fitted to the direct beam data using the runs without target, but
consistent with the RMS of the latter. 

\vspace{0.75cm}

Combining information from these three techniques one is able to map out the 
momentum resolution and scale in both data and Monte Carlo for comparison.  The results indicate 
good agreement across a range of momenta and angles allowing us to utilize
the Monte Carlo simulation to generate the momentum resolution correction matrix,
$M^{-1}_{pp^{'}}(\theta^{'})$.

\section{Particle identification}\label{sec:pid}

The particle identification method used here follows exactly that used in the analysis of the p-Al data \cite{ref:alPaper}.
However, a significant improvement in efficiency and purity of the PID and, consequently, 
a large reduction in the systematic uncertainties on the cross-section have been realized by improving the 
association of PID detector hits with reconstructed tracks. We include here a description of the PID method for completeness 
and clarity and a full explanation of how detector hits are selected follows in the next section. 
 
\subsection{The pion-proton PID estimators and PID efficiency calculation}

The analysis uses particle identification information from the time-of-flight
and Cherenkov PID systems; 
the discrimination power of time-of-flight below 3 \GeVc and the 
Cherenkov detector above 3 \GeVc are combined to provide powerful separation of pions and protons.  The calorimeter
is presently used only for separating pions and electrons when characterizing the response of the other 
detectors. The resulting efficiency and purity of pion identification
in the analysis region is excellent.    

Particle identification is performed by determining the probability that a given track is a
pion or a proton based on the expected response of the detectors to each particle type and the measured
response for the track.  Information from both detectors is combined for maximum discrimination power using a Bayesian technique,
\vspace{0.5ex}
\begin{equation}
  P(\alpha|\beta,N_{\mathrm{pe}},p,\theta) = \frac{P(\beta,N_{\mathrm{pe}}|\alpha,p,\theta) \cdot P(\alpha|p,\theta)}
  {\sum\limits_{i = \pi,\mathrm{p},...}
  P(\beta,N_{\mathrm{pe}}|i,p,\theta) \cdot P(i|p,\theta)} \ .
  \label{eq:completeBayes}
\end{equation}
\vspace{0.5ex}

\noindent where $P(\alpha|\beta,N_{\mathrm{pe}},p,\theta)$ is the probability that a track with reconstructed velocity $\beta$, number of 
associated photo-electrons $N_{\mathrm{pe}}$, and momentum and angle $p$ and $\theta$ is a particle of type $\alpha$.
$P(i|p,\theta)$ is the so-called prior probability for each particle type, $i$, and is a function of $p$ and $\theta$.  
In the Bayesian approach, the priors
represent one's knowledge of the relative particle populations before performing a measurement.  
Finally,  $P(\beta,N_{\mathrm{pe}}|i,p,\theta)$ is the expected response ($\beta$ and $N_{\mathrm{pe}}$) of the PID detectors for a particle 
of type $i$ and momentum and angle $p,\theta$.  

The following simplifications are applied to Eq. \ref{eq:completeBayes}.  First, we will assume no a priori knowledge of the 
underlying pion/proton spectra; that is, the prior distributions will be flat and equal everywhere, $P(i|p,\theta) = 1$ 
for all $p$, $\theta$.  This allows the priors to be dropped from the expression, but the PID estimator no longer has a full 
probabilistic interpretation and cannot be directly used to estimate the particle yields.  One could iterate the probability 
distributions to determine the yields.  Alternatively, one can build the PID estimator for each track independently, and 
an efficiency and migration must be determined for a given cut on the estimator value, $P_{\mathrm{track}}>P_{\mathrm{cut}}$.  We will see that
the necessary corrections are small and the systematic uncertainty is negligible compared to other sources, making this approach
adequate. Second, we will consider the response functions of the different PID detectors as independent and can therefore
factorize the probability into separate terms for the TOFW and CHE.  
Third, with the new detector hit selections, the time-of-flight and Cherenkov detector 
responses show no angular dependence allowing  $\theta$ to be removed from the above expression.  
Finally, we will only consider pions and protons as possible secondary particle types.  Monte Carlo simulation
shows other potential backgrounds to a \piplus yield measurement to be small.
\vspace{0.5ex}
\begin{equation}
  \begin{split}
    P&(\alpha|\beta,N_{\mathrm{pe}},p) = \\
    & \frac{P(\beta|\alpha,p) \cdot P(N_{\mathrm{pe}}|\alpha,p)}
    {P(\beta,|\pi,p)\cdot P(N_{\mathrm{pe}}|\pi,p) + P(\beta|\mathrm{p},p) \cdot P(N_{\mathrm{pe}}|\mathrm{p},p)} \ .
    \label{eq:simpleBayes}
  \end{split}
\end{equation}
\vspace{0.5ex}

\noindent where $P(\alpha|\beta,N_{\mathrm{pe}},p)$ is the PID estimator for a track with reconstructed $\beta$, $N_{\mathrm{pe}}$ and $p$ 
to be of type $\alpha$
 and $P(\beta|\mathrm{p},p)$ and $P(N_{\mathrm{pe}}|\mathrm{p},p)$ are the response functions for the TOFW and CHE, respectively, 
which will be fully described below.
Pions are selected by making a cut in this PID variable,
equal to 0.6 in the present analysis.  

The efficiency for 
pion selection and the migration between pions and protons can be calculated 
analytically from the parametrized detector response functions for a given cut in probability.     
These corrections consist
(for each momentum bin) of a 2x2 matrix (PID efficiency matrix)
\vspace{0.2ex}
\begin{equation}
  \left ( \begin{array}{c} \pi \\
    \mathrm{p} \end{array} \right )_{\mathrm{rec}} = \left ( \begin{array}{cc} M_{\pi\pi} & M_{\pi \mathrm{p}} \\
      M_{\mathrm{p} \pi} & M_{\mathrm{pp}} \end{array} \right ) \cdot  \left (
    \begin{array}{c} \pi \\ 
      \mathrm{p} \end{array} \right )_{\mathrm{true}}  \ ,
  \label{eq:pidEffMatrix}
\end{equation}
\vspace{0.5ex}

\noindent where  $M_{\pi\pi}$ and  $M_{\mathrm{pp}}$ are the efficiencies for
correctly identifying pions and protons, respectively, and  
$M_{\pi \mathrm{p}}$ and $M_{\mathrm{p} \pi}$ are the migration terms of 
 true protons identified as pions and true pions identified as protons,
respectively.
The full covariance matrix (16 terms) of the PID efficiency matrix is also computed analytically. 
This calculation is explained fully in \cite{ref:pidPaper}.  
The final expression for the PID efficiency--migration matrix elements reads:
\vspace{0.5ex}
\begin{equation}
    M_{ij} = \sum_{\ches=0,1} C_j^\ches \cdot \left[\int_{\beta_\pi^\ches,0}^{\infty,\beta_\pr^\ches} d\beta \,\, G_j(\beta) \newline
      + \omega_j^{\pi\pr} \cdot S(P_i^\ches|_{\mathrm{CHE}}>P_{\mathrm{cut}}) \right] \ ,
    \label{eq:eff_matrix}
\end{equation}
\vspace{0.5ex}

\noindent where $C_j^\ches$ is the Cherenkov efficiency (Fig. \ref{fig:ckovPdf}), $G_j(\beta)$ and $\omega_j^{\pi\pr}$ are the Gaussian (Fig. \ref{fig:betaPdf}) and
non-Gaussian contributions to the beta response, respectively, and $S(P_i^\ches|_{\mathrm{CHE}}>P_{\mathrm{cut}})$ is a step function
controlling the integration of the non-Gaussian part, $\omega$.

The 2x2 matrix of Eq. \ref{eq:pidEffMatrix} is easily inverted for converting reconstructed yields into true yields of pions and protons.  The elements of the
matrix in Eq. \ref{eq:pidEffMatrix} are shown in Fig. \ref{fig:pidEff} as a function of particle momentum.  The pion efficiencies are
$>$95\% and the proton-pion migrations are all less than 1\%.  But first we must desribe the detector response functions that are 
a key input to the efficiency and migration calculations.

\subsection{PID detector hit selection and response functions}

This section describes the quality criteria applied to select PID detector hits and the resulting response functions for 
pions and protons.
Previously, PID detector hits have been associated with reconstructed tracks based only on a geometrical matching criterion.
A Kalman filter package is used to extrapolate each track to the plane of each detector, 
and this position is compared to the reconstructed $x,y$ positions of all reconstructed hits
in that detector.  The detector hit with the best matching $\chi^2$ was then assigned to that track. 
In this scheme the reconstruction can associate a single detector hit with multiple
tracks and each track likely has additional candidate detector hits which are being ignored.  In particular, in \cite{ref:alPaper}
it was seen that a fraction of protons had a non-negligible amount of associated photo-electrons due to light from pions
or electrons being wrongly associated with proton tracks. Also in \cite{ref:alPaper} the TOFW response function
contained a non-gaussian component where $\approx$10\% of reconstructed $\beta$ fell greater than 5$\sigma$ from the
mean expectation and could not be used for identification.
Additional criteria have been developed for selecting PID detector hits to address these issues
and have led to a reduction in PID backgrounds by as much as a factor of 10 in some regions of phase space.
This is the source of the drastic improvement in PID systematics since our previous publication (3.5\% to 0.5\%). 
The PID hit selection is described in detail below.

\subsubsection{Time-of-flight response}\label{sec:tofhitselect}

A time-of-flight measurement is required for particle identification in this analysis.  It was discovered, due to the presence of 
a significant, almost flat background far from the Gaussian peaks in the $\beta$ distributions, that a more strict set of selection cuts was
required to ensure a quality time-of-flight measurement.  The small efficiency loss due to this selection can be measured directly from the data and
will be combined with the tracking efficiency discussed above to form an overall reconstruction efficiency,
\vspace{0.5ex}
\begin{equation}
  \varepsilon^{\mathrm{recon}} = \varepsilon^{\mathrm{track}} \cdot \varepsilon^{\mathrm{TOFW-match}} \ .
\end{equation}

\begin{figure}
  \begin{center}
    \includegraphics[width=4.2cm]{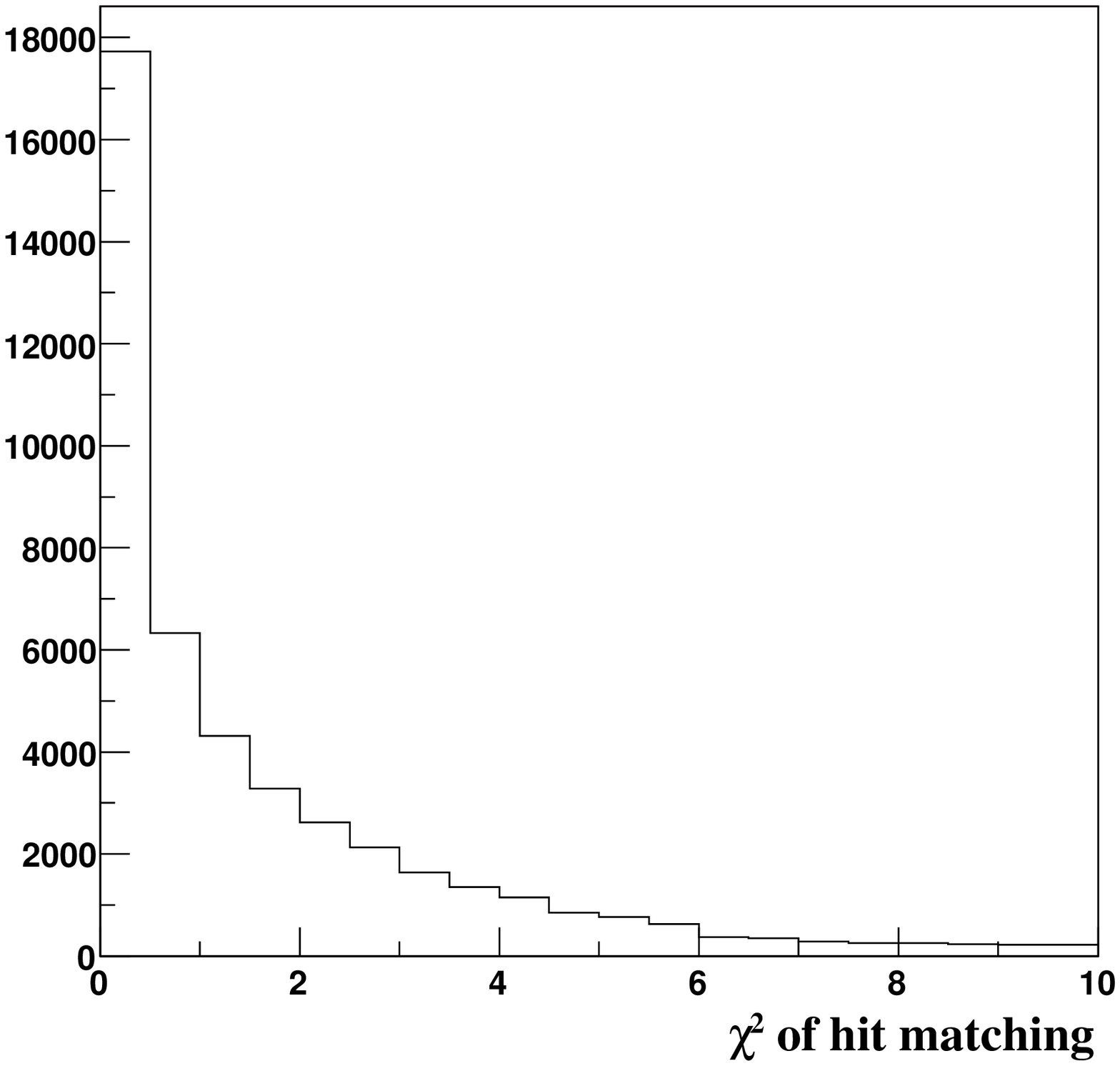}
    \includegraphics[width=4.2cm]{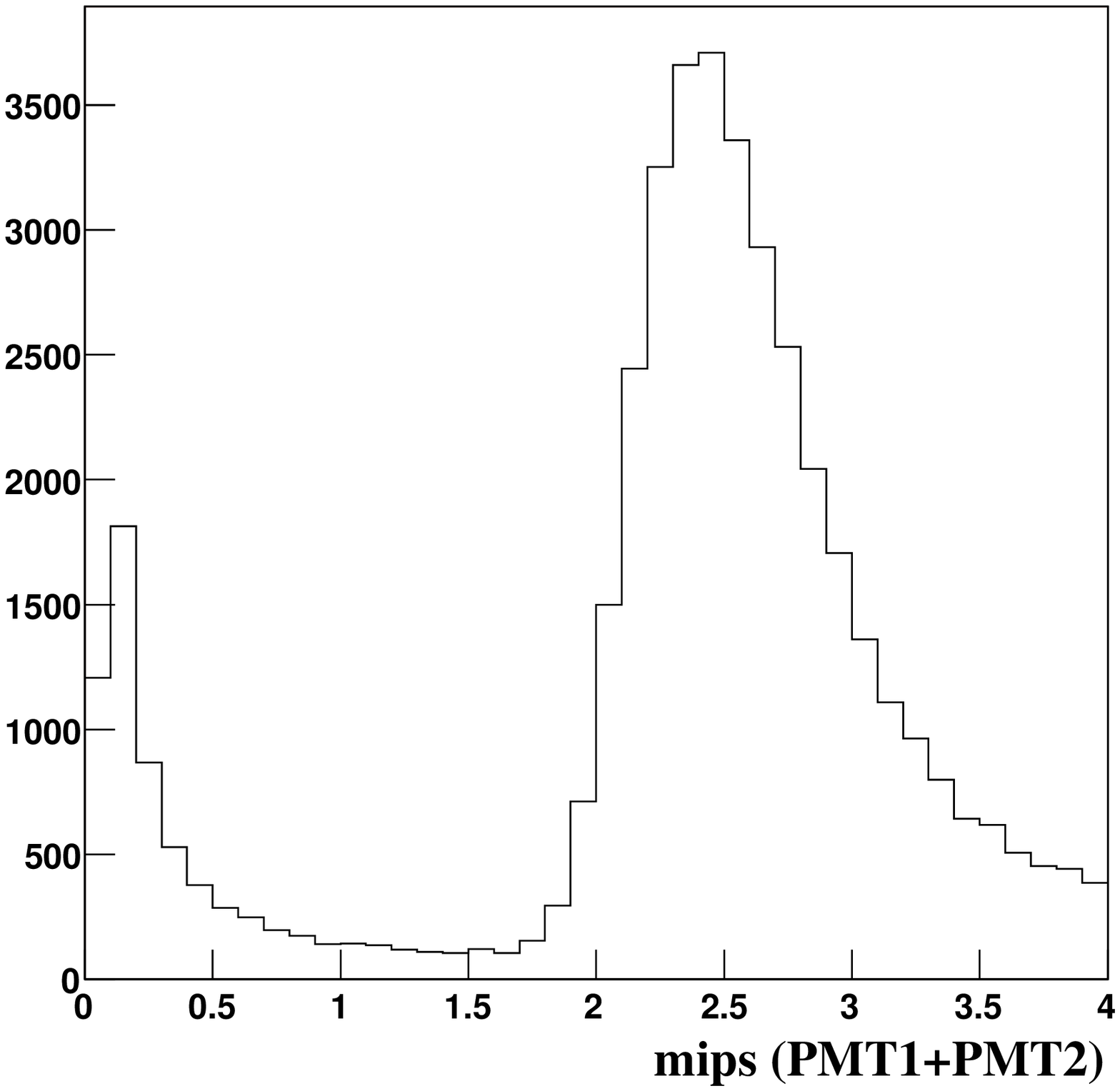}
    \caption{\label{fig:tofSelVars} TOFW hit reconstructed variables.  The left panel shows the distribution of the $\chi^2$
      between the extrapolated track position and the reconstructed scintillator hit position.  The right panel shows the total
      reconstructed number of minimum ionizing particles (mips) from the two PMTs on the scintillator volume that was hit. Time-of-flight hits are selected by
      requiring a $\chi^2 \leq 6$ and number of mips $\geq 1.5$ (see the text).}
  \end{center}
\end{figure}

Each track can have multiple time-of-flight measurements ($TOFW - t_0$) associated with it in the reconstruction.
It is possible for a single hit to match with multiple tracks if the tracks are close enough together when hitting the wall.  It is also possible 
that electromagnetic showers associated with a particle passing through detector material can create additional hits beyond the primary hit caused by 
the hadron of interest.  
To minimize inaccurate time measurements due to these effects, the time-of-flight candidates for each track are time-ordered and the
earliest hit passing the following criteria is selected:
\begin{itemize}
\item if the track shares the TOFW hit with another track, it must have the better geometric matching $\chi^2$;
\item $\chi^2$ of the geometrical matching between the track and TOFW hit $\leq 6$; 
\item total reconstructed number of minimum ionizing particles (mips) from the two PMTs in a hit $\geq 1.5$. 
\end{itemize}
The $\chi^2$ distribution for track-TOFW hit matching and the total pulse-height distribution for this data set
 are shown in Fig. \ref{fig:tofSelVars}.

Having applied these criteria to time-of-flight measurements we must understand the associated efficiency loss as well as the remaining level
of non-Gaussian component to the $\beta$ spectrum.  Each of these have been carefully measured and the needed corrections applied.  

\begin{figure}
  \begin{center}
    \includegraphics[width=9cm,height=5cm,clip=true,trim=0.5cm -0.4cm 0cm 0cm]{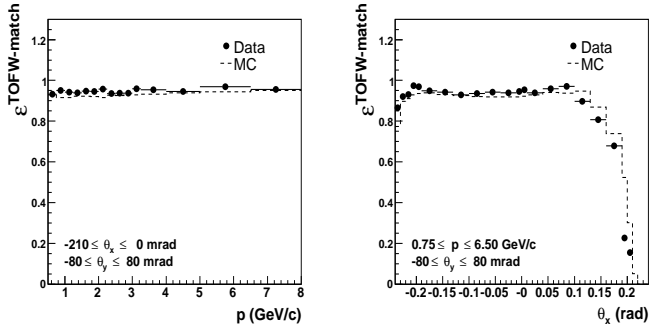}
    \caption{\label{fig:tofMatchEff} TOFW matching efficiency as a function of particle
      momentum (left) and production angle in the horizontal plane, $\theta_x$ (right) as measured from data and Monte Carlo.  
      The TOFW matching efficiency does not have the momentum dependence of the tracking efficiency, but does
      exhibit the same effects of geometric acceptance as the drift chambers as seen in the right, $\theta_x$, plot.
      Note the present analysis is performed using tracks in the range $-0.210 \ \rad \leq \theta_x \leq 0 \ \rad$ where the acceptance is flat in momentum.}
  \end{center}
\end{figure}

The efficiency is measured from the data by using a sample of
reconstructed tracks which leave a signal in the calorimeter
(downstream of the scintillator wall) and asking how often a
time-of-flight measurement passing selection cuts is 
found.  Fig. \ref{fig:tofMatchEff} shows the matching efficiency for
data and Monte Carlo to be flat in momentum and around 95\% in
the data. 
This efficiency is combined with the tracking efficiency to provide the total analysis track 
reconstruction efficiency. 
The efficiencies generated from the data themselves have been 
used in the analysis for the results being presented here.

\begin{figure}[h!]
  \begin{center}
    \includegraphics[width=4.2cm,height=4.2cm,clip=true,trim= 1.0cm 0cm 1cm 0cm]{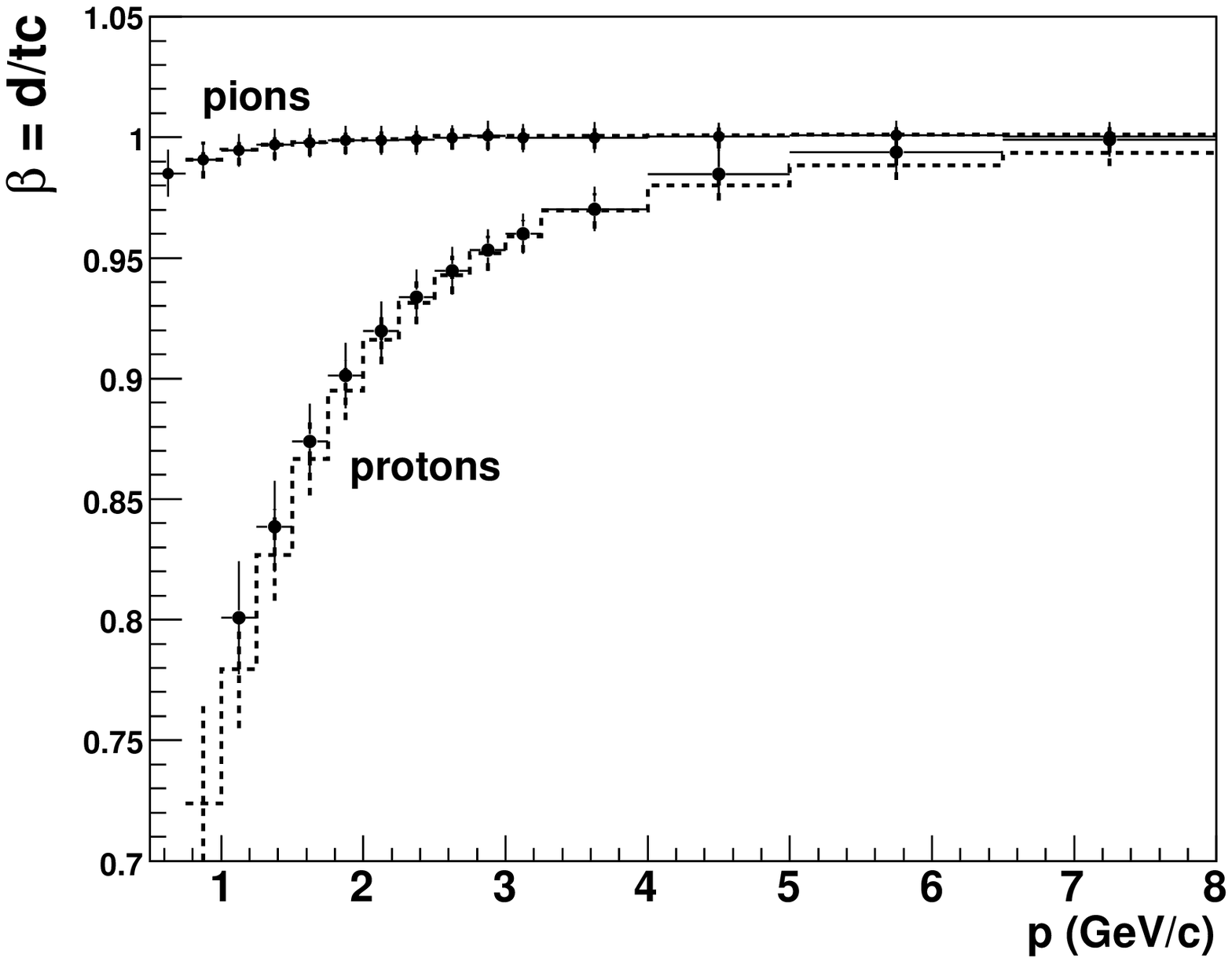}
    \includegraphics[width=4.2cm,height=4.2cm,clip=true,trim= 0.4cm 0cm 1.7cm 0cm]{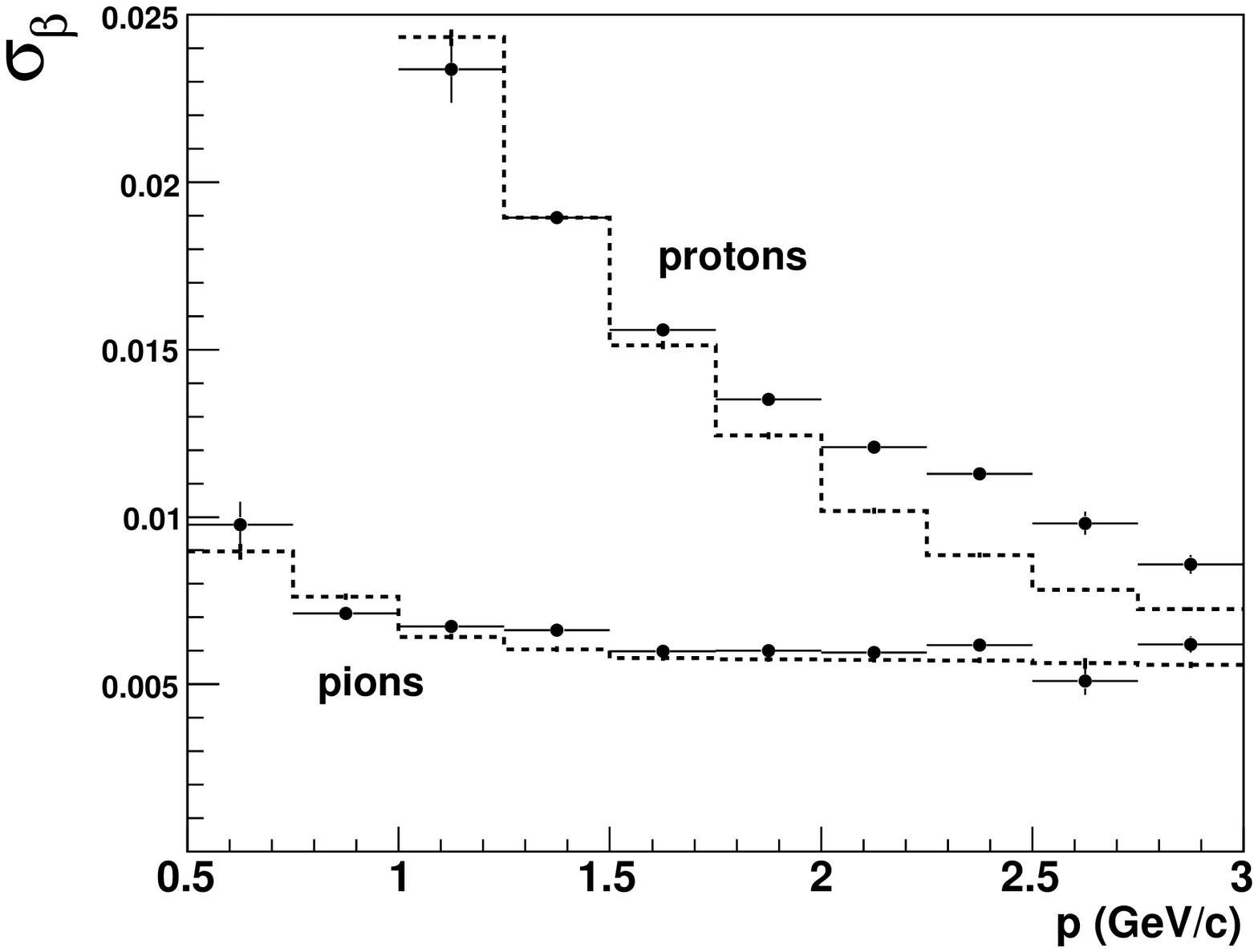}
    \caption{\label{fig:betaPdf}
      $\beta$ response for pions and protons as measured from data and Monte Carlo. The left panel shows the mean beta
      values with the error bars representing the width of the Gaussian distribution.  The right panel highlights the width of the distributions
      and shows how the resolution asymptotically approaches $\approx 0.006$. The solid points are the response as measured from data;
      the dashed histograms are the response as measured from Monte Carlo. }
  \end{center}
\end{figure}

Having applied these selection criteria we must characterize the response of the time-of-flight detector for different particle types.
The $\beta$ response has been parameterized by a Gaussian function.
A method has been developed to extract the parameters of this function directly from the data and has already been
described in \cite{ref:alPaper} and \cite{ref:pidPaper}. The method was applied to both data and Monte Carlo events
and the results are shown in Fig. \ref{fig:betaPdf}.  The figure shows a small bias in
the simulation of the time-of-flight resolution for protons above $\approx 2$~\GeVc, resulting in a particle identification
efficiency bias of about 0.5\% (see Fig. \ref{fig:pidEff}).  To avoid any bias introduced by the 
simulation of the time-of-flight system, the response functions as determined from data    
are used in the analysis of data.

\begin{figure}[h!]
  \begin{center}
    \includegraphics[width=4.2cm,height=4.cm,clip=true,trim= 1.0cm 1.4cm 1cm 0cm]{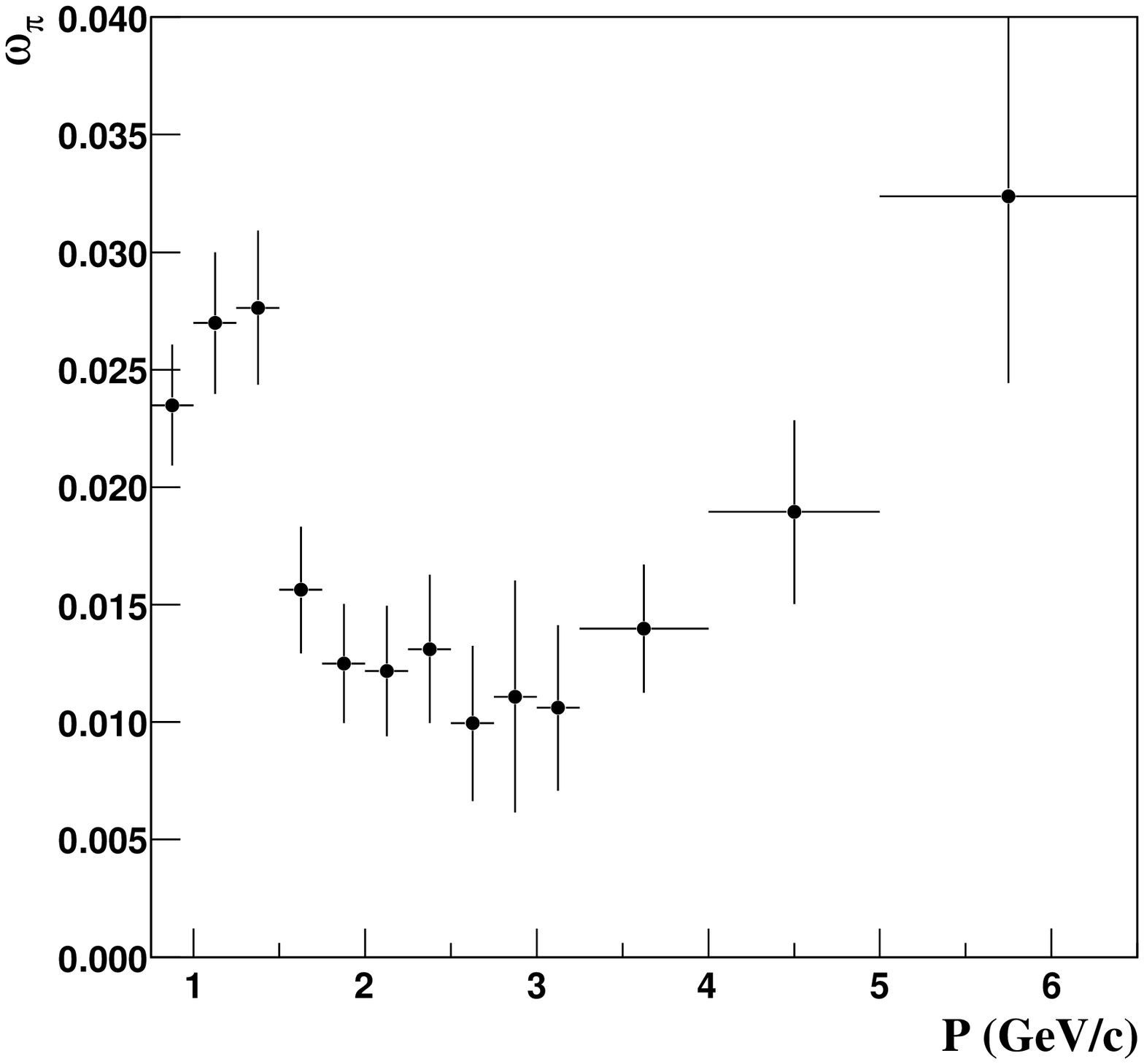}
    \includegraphics[width=4.2cm,height=4.cm,clip=true,trim= 0.4cm 1.4cm 1cm 0cm]{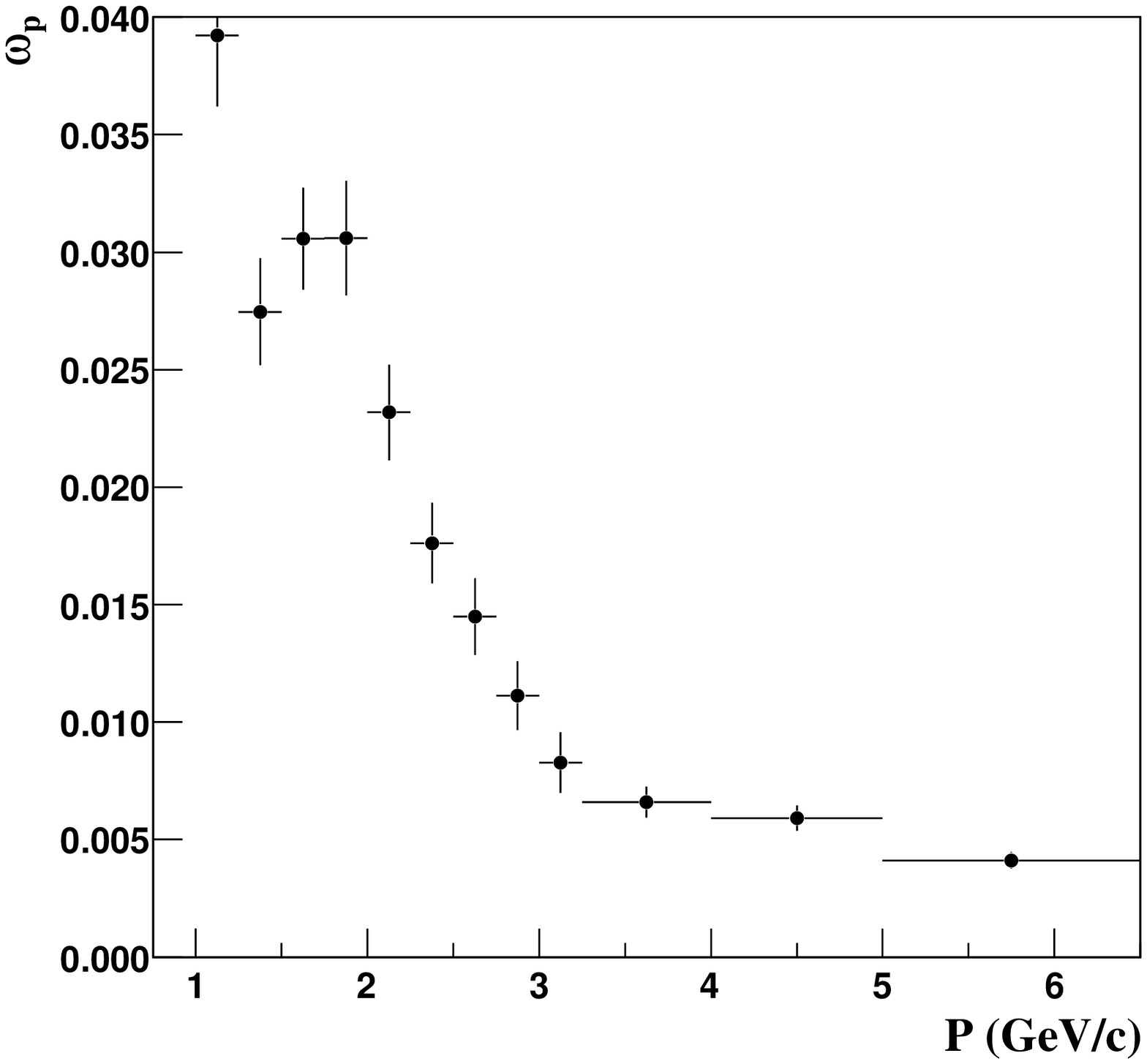}
    \caption{\label{fig:betaOutliers}
      Pion (left) and proton (right) $\beta$-outlier rates as a function of momentum. The pion outlier rate has been measured 
      from the data; the proton outlier rate is estimated using the Monte Carlo. In reference \cite{ref:pidPaper} it has been 
      demonstrated that no bias is seen when using the Monte Carlo to calculate the outlier rate.}  
  \end{center}
\end{figure}

Additionally, there is a small rate of non-Gaussian time-of-flight measurements, 
called ''$\beta$-outliers'', which must be accounted for separately.  $\beta$-outliers are defined as time-of-flight 
measurements greater than $5\sigma$ from the mean of the expected $\beta$ response function.
This small, non-Gaussian component of the time-of-flight response, shown in Fig. \ref{fig:betaOutliers},  
has been fully accounted for in the PID efficiency calculation described above.   
It should be noted that, due to the improvements in the hit selection criteria being described here,
the $\beta$-outlier effect as described in \cite{ref:alPaper} has been reduced from $\sim$10\% to $\sim$1--3\% since that publication.
The outlier rate is the largest contribution to the systematic error coming from particle identification, and with this improvement PID
now makes a negligible contribution to the total systematic error in the cross-section analysis. 

\subsubsection{Cherenkov response}

The Cherenkov detector is used to veto electrons below 3 \GeVc and to differentiate pions from protons above 3 \GeVc.  

\begin{figure}[h!]
  \begin{center}
    \includegraphics[width=4.3cm, trim=1.3cm 1.25cm 0cm 0.5cm]{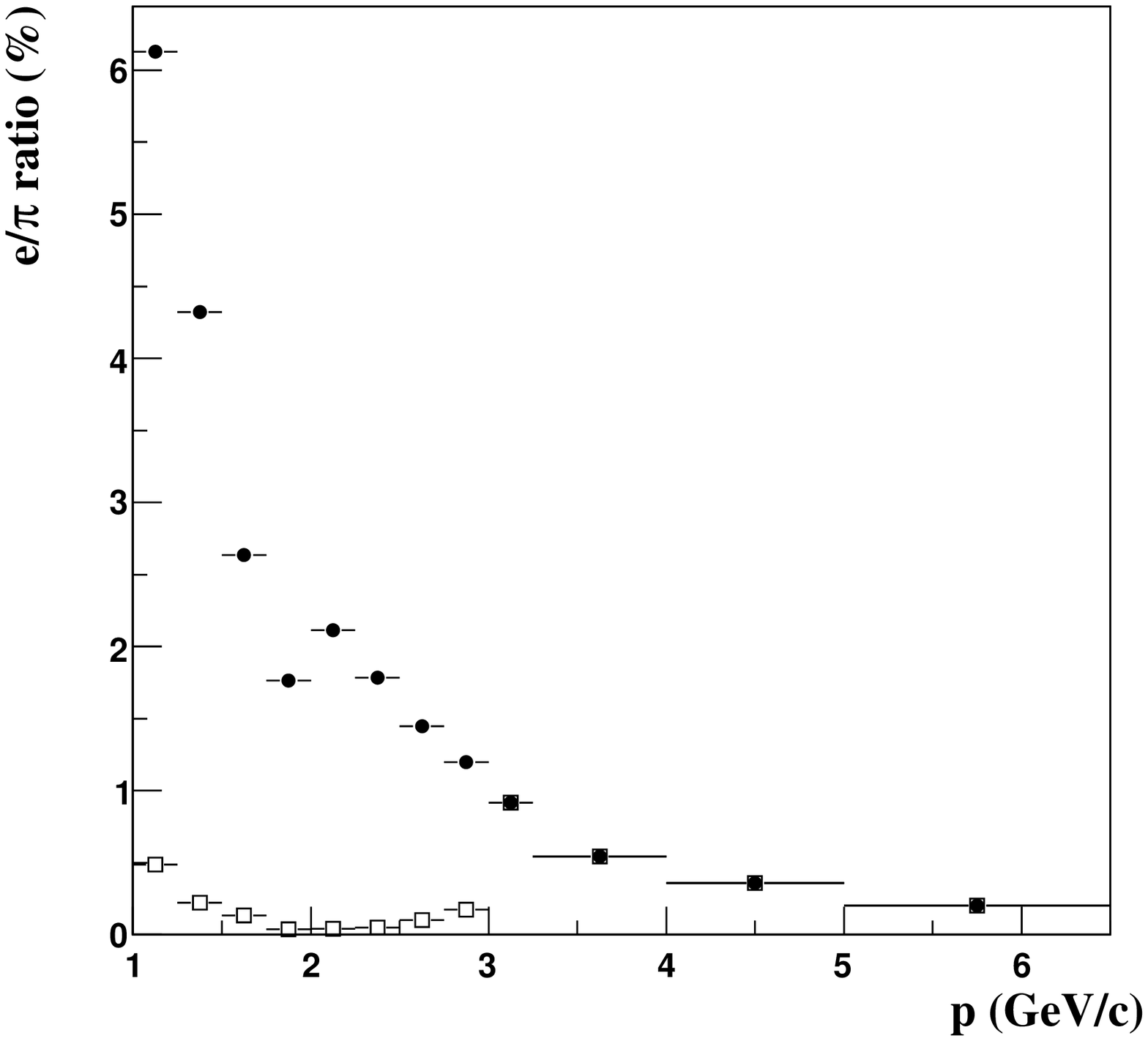}
    \includegraphics[width=4.1cm, trim=0.1cm 0cm 0.3cm 0.4cm]{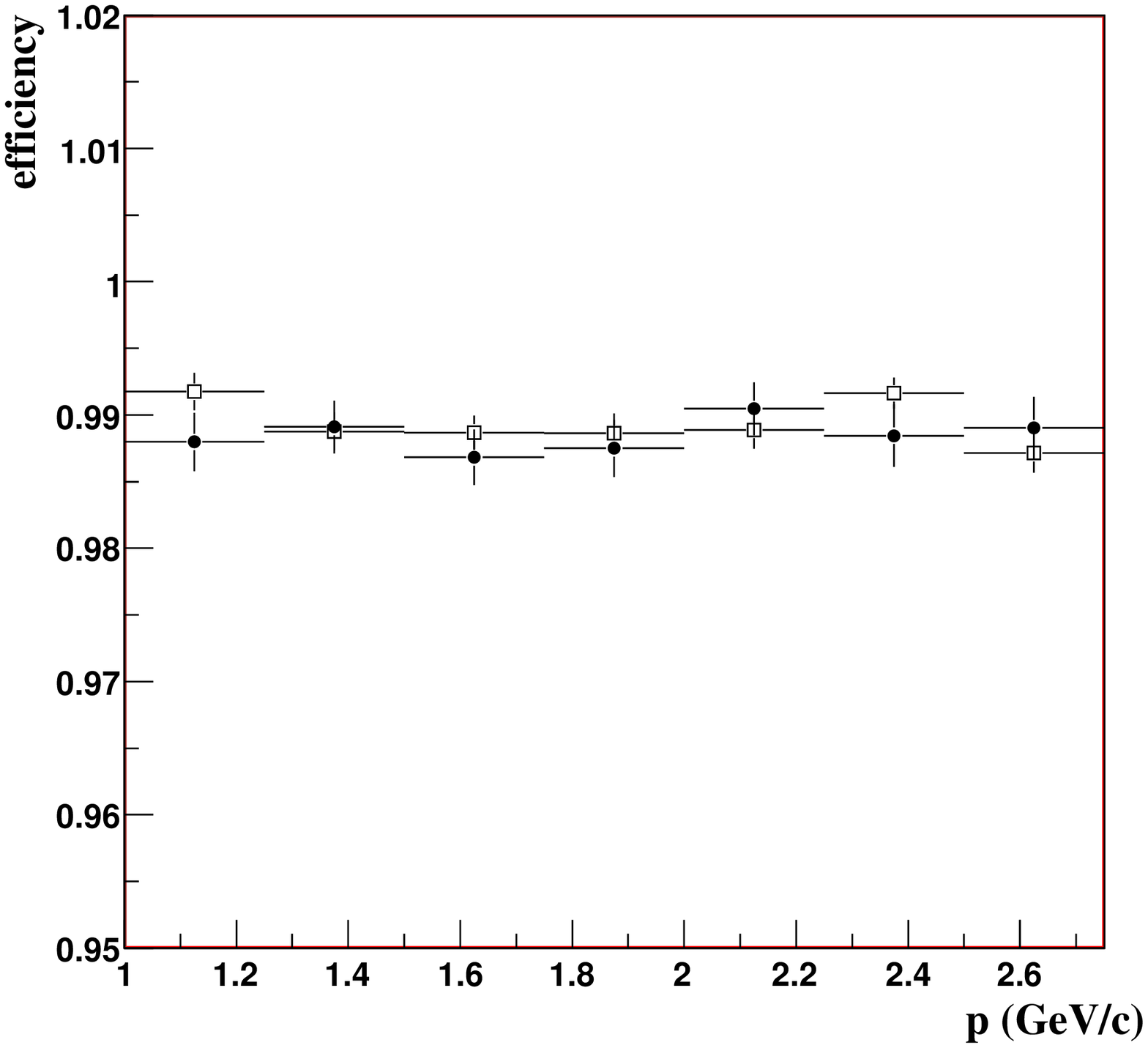}
    \caption{\label{fig:evetoEffect} 
      The left panel shows the $e/\pi$ ratio from a Monte Carlo simulation 
      before (solid points) and after (open squares) the application of a 15 photo-electron cut.  
      This cut reduces the electron contamination to 0.5\% or less in the region where it is applied. The right panel shows the
      efficiency for pions (solid points) and protons (open squares) to pass the 15 photo-electron cut below 3 \GeVc, 
      and is $\approx 99\%$ for both. }
  \end{center}
\end{figure}

Below 3 \GeVc the Cherenkov signal is not used in the calculation of the particle identification probability according to Eq. \ref{eq:simpleBayes}, 
but instead electrons are removed by a simple veto of tracks with greater than 15 photo-electrons.  Fig. \ref{fig:evetoEffect} 
demonstrates the effect of the electron veto cut.  The left panel shows the $e/\pi$ ratio in the Monte Carlo before and after
applying the 15 photo-electron cut below 3 \GeVc.  The remaining electron contamination is less than 1\% everywhere, and less than 0.5\% in the region
where the veto is applied.  One expects a very small efficiency loss for pions and protons due to this cut in photo-electrons and this is also shown in
Fig. \ref{fig:evetoEffect}.  Approximately 1\% of pions and protons do not pass the electron veto; a correction has been applied in the present analysis.

\begin{figure}[h!]
  \begin{center}
    \includegraphics[width=9cm]{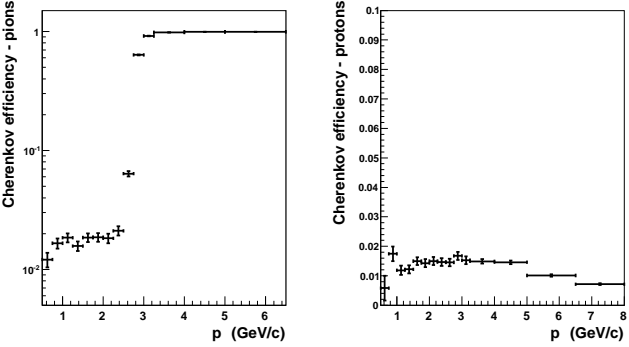}
    \caption{\label{fig:ckovPdf} Cherenkov response for pions (left) and protons (right). The points are the efficiencies for a track to 
      have an associated Cherenkov hit with greater than 2 photo-electrons.  The threshold for pions at around 2.6 \GeVc 
      is clearly visible (note the log scale).  The small efficiency for protons and
      below threshold pions of around 1.5\% is due to false associations with light generated by other particles in the event. }
  \end{center}
\end{figure}

Above 3 \GeVc the Cherenkov is a powerful discriminator of pions and protons. (Monte Carlo simulations indicate that there are a negligible
number of electrons above 3 \GeVc and these are thus ignored.)  Presently the Cherenkov is being used digitally.  That is the spectral information
of the light output is not being used.  Instead we define a signal as an associated hit with greater than 2 photo-electrons.  Two or less is considered 
no signal. Based on this definition we determine the efficiency for pions and protons to have a signal in the Cherenkov as a function of particle momentum.
 Fig. \ref{fig:ckovPdf} shows the expected response for pions and protons in the Cherenkov
both above and below the pion threshold.  Above threshold the Cherenkov is greater than 99\% efficient for pions.   The small efficiency for protons and
pions below threshold of around 1.5\% is due to false associations with light generated by other particles in the event. 

\vspace{0.75cm}

\begin{figure}
  \begin{center}
    \includegraphics[width=4.2cm]{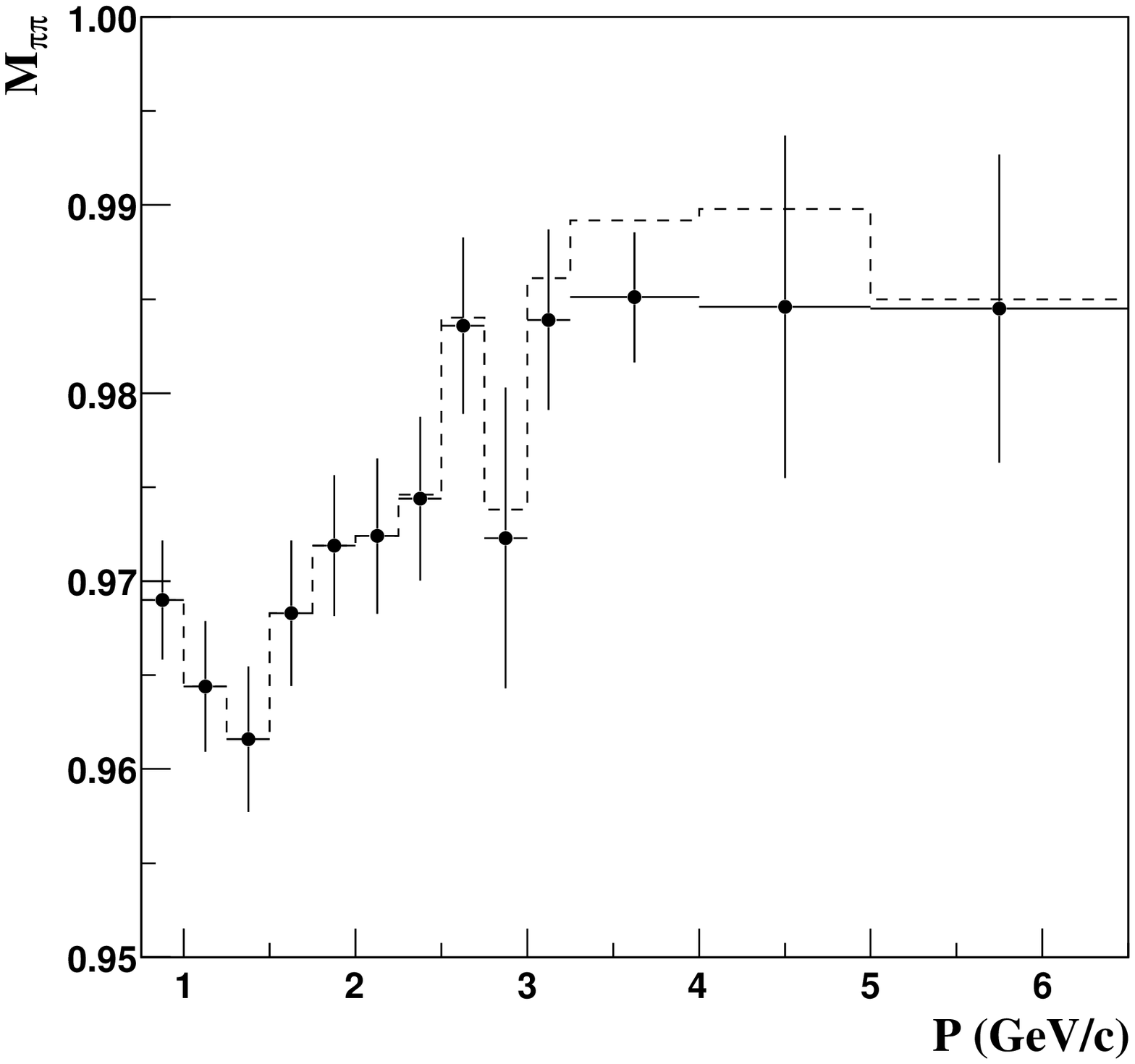}
    \includegraphics[width=4.2cm]{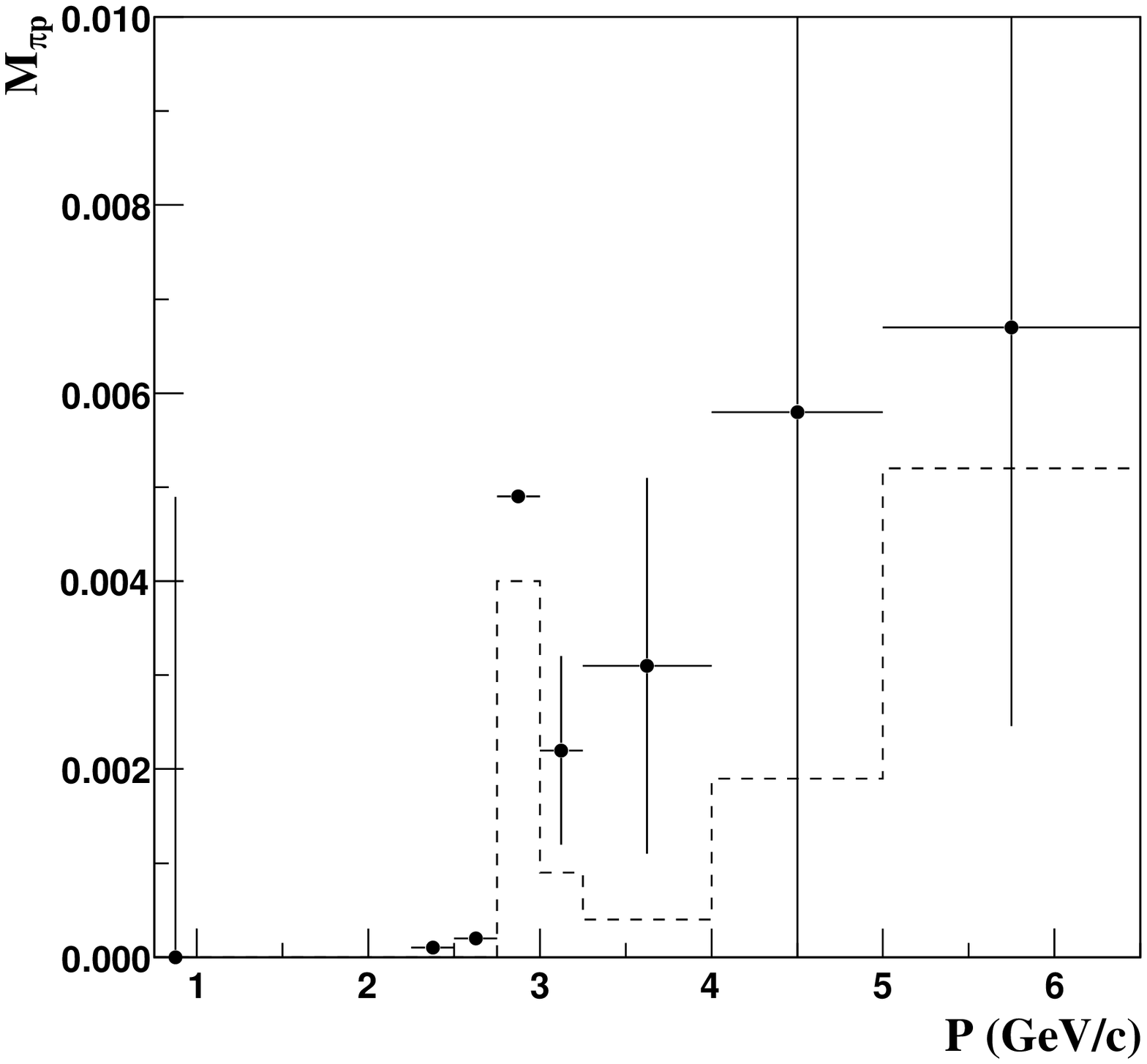}
    \includegraphics[width=4.2cm]{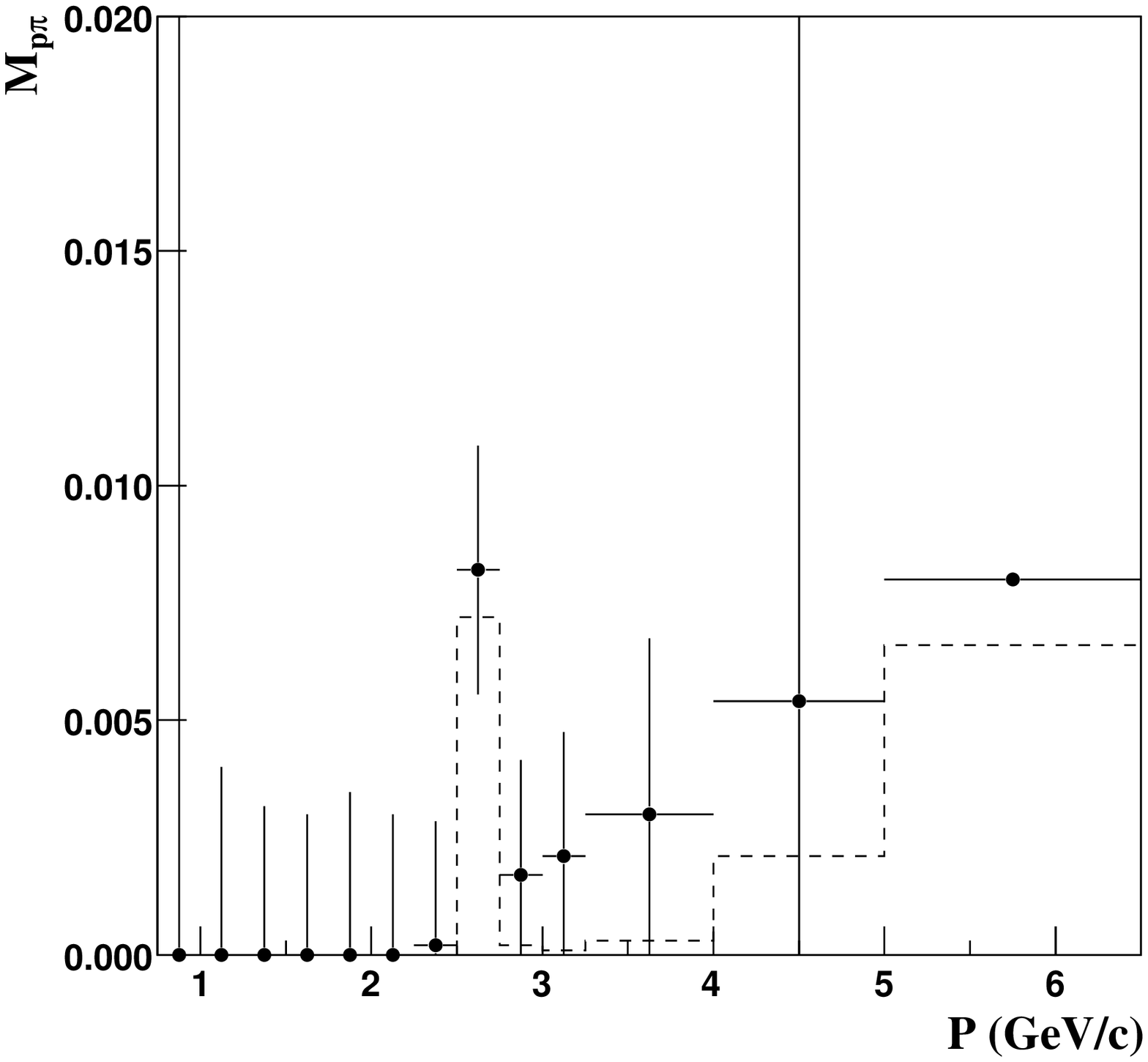}
    \includegraphics[width=4.2cm]{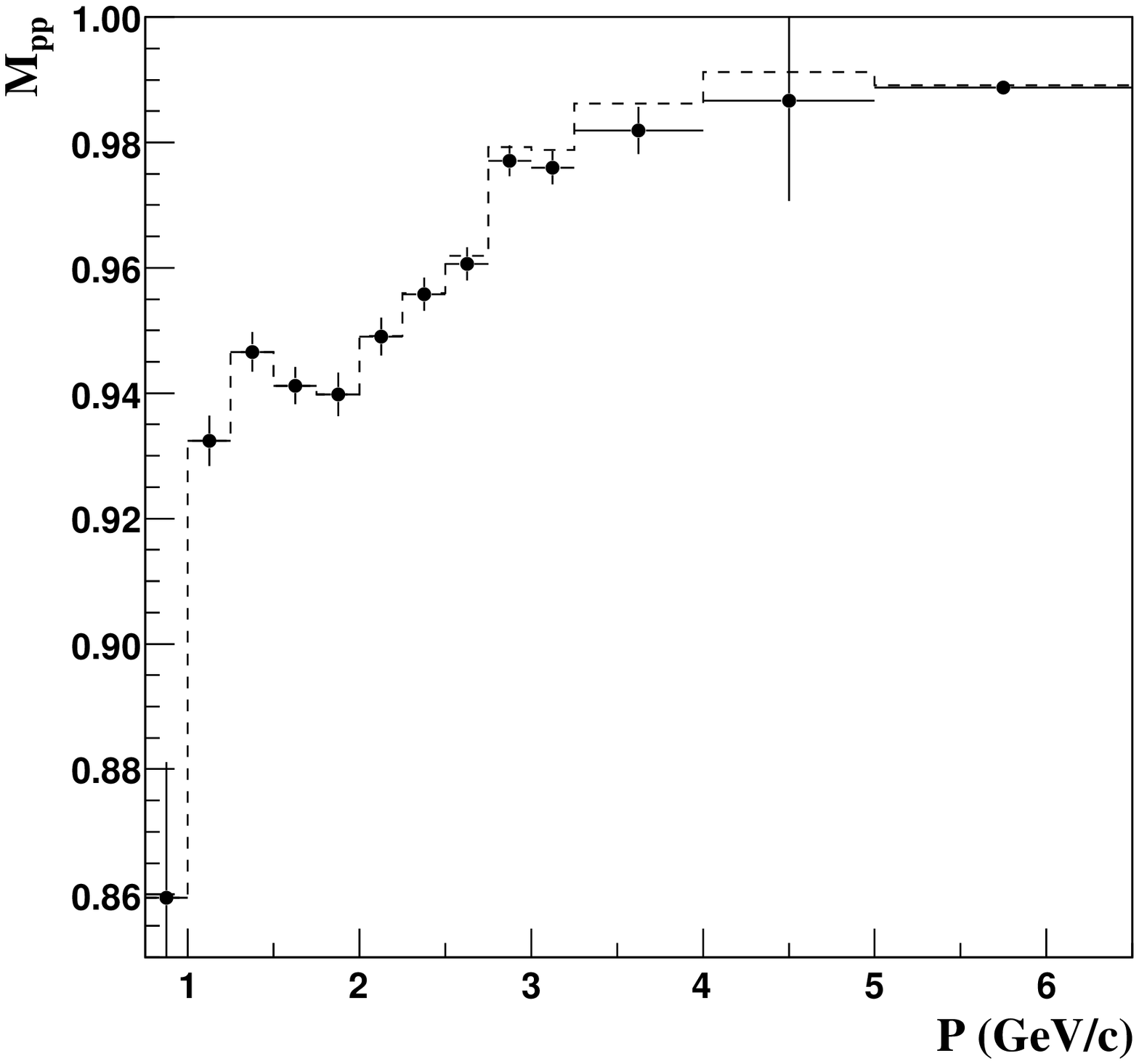}
    \caption{\label{fig:pidEff} Particle ID efficiency and migration matrix elements as a function of momentum.  Upper left is
    the pion identification efficiency, $M_{\pi\pi}$.  Upper right is the proton to pion migration, $M_{\pi \mathrm{p}}$.  Lower left is
    the pion to proton migration, $M_{\mathrm{p} \pi}$.  Lower right is the proton identification efficiency,  $M_{\mathrm{pp}}$.
    Points with errors are the values calculated from data.  The dashed histograms are those determined from the Monte Carlo. 
    The slight data-Monte Carlo bias seen above 2.5 \GeVc results from the bias seen in the TOFW simulation. The values
    determined from data have been used in the analysis to avoid sensitivity to this bias.}
  \end{center}
\end{figure}

Using the characterized responses of the TOFW and CHE detectors we can calculate the PID estimator for reconstructed tracks given in Eq. 
\ref{eq:simpleBayes} and the efficiency-migration matrix elements given by Eq. \ref{eq:eff_matrix} and shown in Fig. \ref{fig:pidEff}.

\section{Physics results}\label{sec:results}

Applying corrections to the raw yields as described in the previous sections and according to
Eq. \ref{eq:finalxsec}, we have calculated the double-differential inelastic cross-section for the
production of positive pions from collisions of 8.9 \GeVc protons with beryllium in the kinematic
range from $0.75 \ \GeVc \leq p_{\pi} \leq 6.5$ \GeVc and $0.030 \
\rad \leq \theta_{\pi} \leq 0.210 \ \rad$.  

\begin{figure}
  \begin{center}
    \includegraphics[width=8cm,clip=true,trim=1cm 0cm 0cm 0cm]{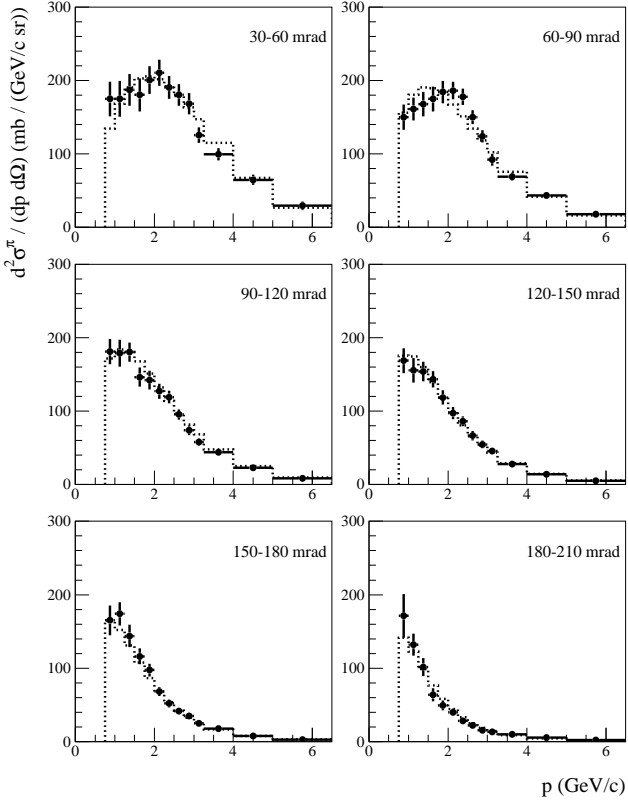}
    \caption{\label{fig:xsecBins}
      Measurement of the double-differential production cross-section of positive pions, $d^2\sigma^{\pi^{+}}/dpd\Omega$,
      from 8.9 \GeVc protons on beryllium as a function of pion momentum, $p$, in bins of pion angle, $\theta$, in the 
      laboratory frame.  The error bars shown include statistical errors and all (diagonal) systematic errors. 
      The dotted histograms show the extended Sanford-Wang parametrization of Eq. \ref{eq:swformula3} 
      with parameter values given in Table \ref{tab:swpar_values_errors}.}
  \end{center}
\end{figure}

\begin{figure}
  \begin{center}
    \includegraphics[width=8cm,clip=true,trim=1cm 0cm 0cm 0cm]{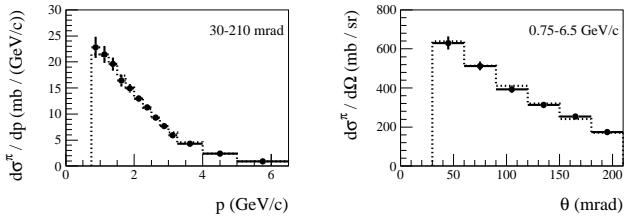}
    \caption{\label{fig:xsec1D}
      Projections of the double-differential cross-section results onto the momentum axis integrated
      over the angular range $30 \ \mrad \leq \theta < 210$ mrad (left) and onto the angle axis integrated
      over the momentum range $0.75 \ \GeVc \leq p < 6.5$ \GeVc (right). Projections for the best-fit extended 
      Sanford-Wang parametrization are also shown, as indicated by dotted histograms.}
  \end{center}
\end{figure}

Systematic errors have been estimated and will be described below.  
A full $(13 \times 6)^2 = 6048$ element covariance matrix has been generated
to describe the correlation among bins. The data are presented graphically as 
a function of momentum in 30 mrad angle bins in Fig. \ref{fig:xsecBins} and 1D projections onto the 
momentum and angle axes are shown in Fig. \ref{fig:xsec1D}.  The central values 
and square-root of the diagonal elements of the covariance matrix are listed in 
Table \ref{tab:xsec_results2}.

\subsection{Error estimation}
\label{sec:errors}

A full systematic error evaluation has been performed on these data in order to estimate the accuracy
of the measurement being presented.  Statistical errors from both the beryllium target data and 
the 8.9 \GeVc empty target data set used to subtract non-target backgrounds
are also included.

The uncertainties associated with the various corrections applied 
have been estimated through a combination of analytical and Monte Carlo techniques.  The approach used here has 
largely followed the methods of \cite{ref:alPaper}.  There are eight sources of systematic uncertainty considered 
which can be grouped into three basic categories: track yield corrections, PID, and momentum reconstruction. 
   
Only the PID uncertainties have been calculated analytically. The covariance matrices of the PID 
efficiency-migration matrices described in Sec. \ref{sec:pid} have been calculated and the errors propagated. 
The cross-section uncertainties from the other sources are estimated by performing
the cross-section calculation N times for N variations of each correction applied.  
The fully correlated error matrix for each correction is then built from the N cross-section results, 
\begin{equation}
  E_{ij}^{\alpha} = \frac{1}{N}\sum_{n=1}^{N}\left[\frac{d^2\sigma^{\pi}_{\mathrm{CV}}}{dpd\Omega} - \frac{d^2\sigma^{\pi}_{\alpha,n}}{dpd\Omega}\right]_i
  \times \left[\frac{d^2\sigma^{\pi}_{\mathrm{CV}}}{dpd\Omega} - \frac{d^2\sigma^{\pi}_{\alpha,n}}{dpd\Omega}\right]_j
\end{equation}
where \emph{i} and \emph{j} label bins of ($p,\theta$), 
$E_{ij}^{\alpha}$ is the $i,j^{th}$ element of one of the error matrices (labeled $\alpha$),
$d^2\sigma^{\pi}_{\mathrm{CV}}/(dpd\Omega)$ is the central value for the
double-differential cross-section measurement and
$d^2\sigma^{\pi}_{\alpha,n}/(dpd\Omega)$ is the cross-section result from the $n^{th}$ variation for the $\alpha^{th}$
systematic.  For example, to estimate the
cross-section uncertainty arising from the absorption correction, 100 analyses are performed where only the
absorption correction is randomly fluctuated 100 times with an RMS of 10\%.  This 10\% is the uncertainty of
the absorption correction as described in Sec. \ref{sec:absorption}.  The total error matrix is just the sum of
the $\alpha$ matrices, $E_{ij} = \sum_{\alpha}E_{ij}^{\alpha}$.

\begin{figure}[h!]
  \begin{center}
    \includegraphics[width=8cm,clip=true,trim=1cm 0cm 0cm 0cm]{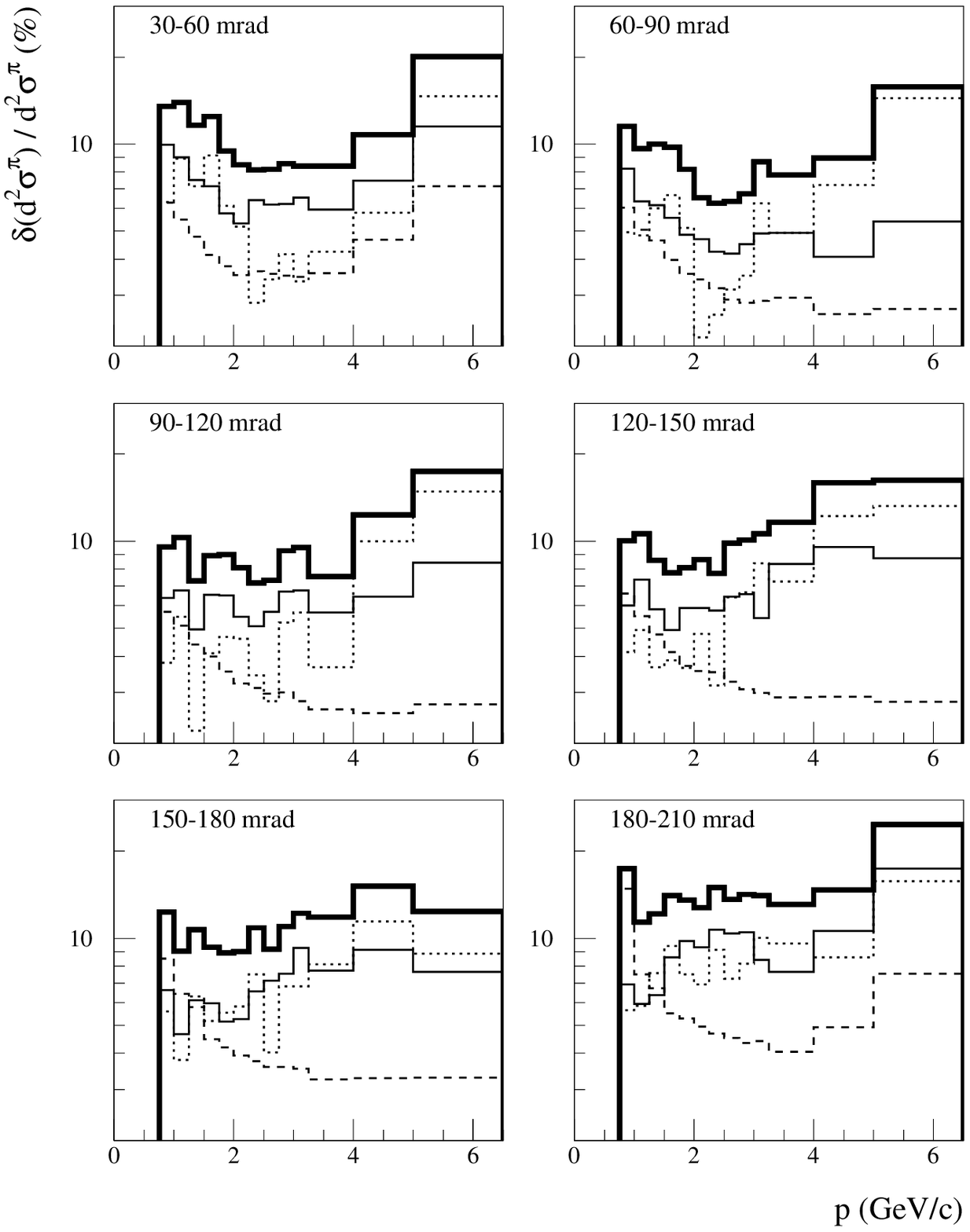}
    \caption{\label{fig:errorsummary}
      Fractional uncertainty (in percent) on the double-differential pion production cross-section measured, as a function of pion momentum and angle. The total uncertainty is shown by the thick black histograms, and individual contributions from the error categories given in Tab.~\ref{tab:errorsummary_delta} are also shown. Statistical, track yield corrections, and momentum reconstruction uncertainties are shown as thin solid, dashed, and dotted histograms, respectively; the overall normalization uncertainty is not shown, and the particle identification uncertainty contribution lies below 2\% for all pion momenta and angles.}
  \end{center}
\end{figure}

The full $78 \times 78$ elements of the covariance matrix will not be published here, but 
to characterize the uncertainties on this measurement we show the square-root of the diagonal
elements of the covariance matrix plotted on the data points in Fig. \ref{fig:xsecBins}.  
Fig. \ref{fig:errorsummary} shows the fractional uncertainty (again diagonal) for each ($p,\theta$) bin.  The total error as well as
the contributions from statistical errors, track yield corrections and momentum reconstruction are shown.
Additionally, we define a 
dimensionless quantity, $\delta_{\mathrm{diff}}$, expressing the typical diagonal error on the double-differential cross-section
\vspace{0.3cm}
\begin{equation}
  \delta_{\mathrm{\small {diff}}}\equiv 
  \frac{\sum_i E_{ii}}{\sum_i (d^2\sigma^{\pi}_{\mathrm{CV}}/(dpd\Omega))_i}
  \label{eq:deltadiff}
\end{equation}
We also define the fractional error on the total integrated pion cross-section in the
range of the measurement $(0.75 \ \GeVc \le p<6.5 \ \GeVc$, $30 \ \mrad
\le \theta<210 \ \mrad)$, $\delta_{\mathrm{int}}$: 
\vspace{0.3cm}
\begin{equation}
  \delta_{\mathrm{\small {{int}}}}\equiv 
  \frac{
    \sqrt{\sum_{i,j}(dpd\Omega)_i
      E_{ij}
      (dpd\Omega)_j
    }}{
    \sum_i (d^2\sigma^{\pi})_i} \ ,
  \label{eq:deltaint}
\end{equation}
where $(d^2\sigma^{\pi})_i$ is the double-differential
cross-section in bin $i$, $(d^2\sigma^{\pi}/(dpd\Omega))_i$,
multiplied by its corresponding phase space element
$(dpd\Omega)_i$. $E_{ij}$ is the covariance matrix
evaluated for the double-differential cross-section data.

For example, the 30\%--40\% absorption correction (see Fig. \ref{fig:absorption}) with a 10\% uncertainty results in an average diagonal 
error of 3.6\% on the cross-section, as one expects.  The uncertainty on the integrated cross-section is approximately 
the same since this correction is a fully correlated yield adjustment.

Table \ref{tab:errorsummary_delta} summarizes these quantities for each of the various
 error sources considered with a typical total
uncertainty of 9.8\% on the double-differential cross-section values being reported 
and a 4.9\% uncertainty on the total integrated cross-section.    
\begin{table*}
  \vspace{3ex}
  \centerline{
    
\begin{tabular}{| l c | c | c |} \hline

{\bf Error Category} & {\bf Comment on method for estimating error} & $\delta_{\mathrm{\small diff}}^{\pi}$ (\%) & $\delta_{\mathrm{\small int}}^{\pi}$ (\%) \\ \hline
Statistical Errors: & & & \\    
\hspace{0.5cm}{\small Be target statistics} & {\small statistical error} & 4.2 & 0.6 \\
\hspace{0.5cm}{\small Empty target subtraction} & {\small statistical error} & 4.6 & 0.6 \\ 
\hspace{0.5cm}{\small \bf Sub-total} &   & {\bf 6.3} & {\bf 0.8} \\ \hline

Track yield corrections: & & & \\
\hspace{0.5cm}{\small Reconstruction efficiency} & {\small stat of tracking efficiency computation sample} & 1.3 & 0.8 \\
\hspace{0.5cm}{\small Pion, proton absorption} & {\small 10\% uncertainty on $\pi$, p absorption rates} & 3.6 & 3.7 \\
\hspace{0.5cm}{\small Tertiary subtraction} & {\small 50\% uncertainty on tertiary production rate} & 1.8 & 1.8 \\
\hspace{0.5cm}{\small Empty target subtraction} & {\small 5\% uncertainty on empty target subtraction normalization} & 1.3 & 1.2 \\
\hspace{0.5cm}{\small \bf Sub-total} &  & {\bf 4.6} & {\bf 4.3} \\ \hline

Particle Identification: & & & \\
\hspace{0.5cm}{\small Electron veto} & {\small stat of electron veto efficiency computation sample} & 0.2 & $<$0.1 \\
\hspace{0.5cm}{\small Pion, proton ID correction} & {\small analytical propagation of errors from parameterized} & 0.4 & 0.1 \\ 
\hspace{0.5cm}{\small \bf Sub-total} &  \raisebox{3ex}{\small PID detector response functions as described in \cite{ref:pidPaper}} & {\bf 0.5} & {\bf 0.1} \\ \hline

Momentum reconstruction: & & & \\
\hspace{0.5cm}{\small Momentum scale} &  {\small 2\% uncertainty on absolute momentum scale} & 3.6 & 0.1 \\
\hspace{0.5cm}{\small Momentum resolution} & {\small different hadronic generators used to generate correction} & 3.4 & 1.0 \\
\hspace{0.5cm}{\small \bf Sub-total} &  & {\bf 5.2} & {\bf 1.0} \\ \hline

Overall normalization: & {\small targeting eff., fully correlated recon. and PID contributions} & {\bf 2.0} & {\bf 2.0} \\ \hline

{\bf Total} & & {\normalsize \bf 9.8} & {\normalsize \bf 4.9} \\ \hline 

\end{tabular}

  }
  \vspace{3ex}
  \caption{\label{tab:errorsummary_delta}
    Summary of the uncertainties affecting the double-differential
    cross-section ($\delta_{\mathrm{\small {{diff}}}}$) and
    integrated cross-section ($\delta_{\mathrm{\small {{int}}}}$) measurements (defined in text).
  }
  \vspace{3ex}
\end{table*}

\subsection{Parametrization of pion production data}\label{sec:sw-param}

Sanford and Wang \cite{ref:SW} have developed an empirical parametrization for describing the production cross-sections of mesons in proton-nucleus interactions. This parametrization has the functional form: 
\vspace{0.3cm}
\begin{equation}
\label{eq:swformula}
\frac{d^2\sigma (\hbox{p+A}\rightarrow \pi^++X)}{dpd\Omega}(p,\theta) =
 \exp[A]p^{c_{2}}(1-\frac{p}{p_{\hbox{\footnotesize beam}}}) \ ,
\end{equation} 
where:
\vspace{0.3cm}
\begin{equation}
\label{eq:swformula2}
A=c_{1}-c_{3}\frac{p^{c_{4}}}{p_{\hbox{\footnotesize beam}}^{c_{5}}}-c_{6}
 \theta (p-c_{7} p_{\hbox{\footnotesize {beam}}} \cos^{c_{8}}\theta ) \ ,
\end{equation} 
and $X$ denotes any system of other particles in the final state, $p_{\hbox{\footnotesize {beam}}}$ is the proton beam momentum in \GeVc, $p$ and $\theta$ are the $\pi^+$ momentum and angle in units of \GeVc and radians, respectively, $d^2\sigma/(dpd\Omega)$ is expressed in units of mb/(\GeVc)/sr, $d\Omega\equiv 2\pi\ d(\cos\theta )$, and the parameters $c_1,\ldots ,c_8$ are obtained from fits to meson production data. 

The parameter $c_1$ is an overall normalization factor, the four parameters $c_2,c_3,c_4,c_5$ describe the momentum distribution of the secondary pions in the forward direction, and the three parameters $c_6,c_7,c_8$ describe the corrections to the pion momentum distribution for pion production angles that are different from zero.

The \piplus production data reported here have been fitted to the empirical Sanford-Wang formula. In the $\chi^2$ minimization procedure, seven out of these eight parameters were allowed to vary. The parameter $c_5$ was fixed to the conventional value $c_5\equiv c_4$, since the cross-section dependence on the proton beam momentum cannot be addressed by the present HARP data-set, which includes exclusively measurements taken at $p_{\hbox{\footnotesize {{beam}}}}=8.9$~\GeVc. In the $\chi^2$ minimization, the full error matrix was used. The goodness-of-fit of the Sanford-Wang parametrization hypothesis for the HARP results can be assessed by considering the best-fit $\chi^2$ value of $\chi^2_{\hbox{\footnotesize {{min}}}}=248$ for 71 degrees of freedom, indicating a very poor fit quality. In particular, inspection of the HARP inclusive pion production double-differential cross-section, and resulting Sanford-Wang parametrization, \\ points to a description of the ratio $g(\theta)$ of the pion momentum distribution at $\theta\neq 0$ with respect to the $\theta = 0$ pion momentum distribution that is more complicated than what can be accommodated within the Sanford-Wang formula, where this ratio is given by $g(\theta)=\exp[-c_6\theta (p-p_c)]$, with $p_c\equiv c_{7} p_{\hbox{\footnotesize {beam}}} \cos^{c_{8}}\theta$.

Given the poor description of this HARP pion production data-set in terms of the original Sanford-Wang parametrization, we explored alternative functional forms. We found a significantly better representation of the data by adopting a simple generalization of the Sanford-Wang formula, obtained by introducing one extra-parameter $c_9\neq 0$ for the description of the angular dependence of the pion momentum distribution, according to $g(\theta)=(1+p/p_{\hbox{\footnotesize {beam}}})^{c_9\theta(p-p_c)}\exp[-c_6\theta (p-p_c)]$. Overall, we use the following parametrization for the inclusive $\piplus$ production double-differential cross-section:
\vspace{0.3cm}
\begin{equation}
  \begin{split}
    \frac{d^2\sigma}{dpd\Omega}(p,\theta) = &
     \exp[A]p^{c_{2}}(1-\frac{p}{p_{\hbox{\footnotesize beam}}}) \\
    & \times (1+\frac{p}{p_{\hbox{\footnotesize beam}}})^{c_9\theta(p-c_{7} p_{\hbox{\footnotesize {beam}}} \cos^{c_{8}}\theta )} \ ,
    \label{eq:swformula3}
  \end{split}
\end{equation}   
\noindent where the argument $A$ in the exponent is given by Eq.~\ref{eq:swformula2}. We obtain in this case a best-fit $\chi^2$ value of $\chi^2_{\hbox{\footnotesize {{min}}}}=117$ for 70 degrees of freedom. 

Concerning the parameters estimation, the best-fit values of the extended Sanford-Wang parameter set discussed above are reported in Table~\ref{tab:swpar_values_errors}, together with their errors. The fit parameter errors are estimated
by requiring $\Delta\chi^2\equiv \chi^2-\chi^2_{\hbox{\footnotesize {{min}}}} = 9.30$, corresponding to the 68.27\% confidence level region for eight variable parameters. Significant correlations among fit parameters are found, as shown by the correlation matrix given in Table~\ref{tab:swpar_correlations}.
\begin{table*}
\centerline{
\begin{tabular}{|c|c|} \hline
{\bf Parameter} & {\bf Value} \\ \hline
$c_1$      & $(5.13\pm 0.41)$ \\
$c_2$      & $(1.87\pm 0.52)$ \\
$c_3$      & $(6.67\pm 1.69)$ \\
$c_4=c_5$  & $(1.56\pm 0.55)$ \\
$c_6$      & $(1.19\pm 0.18)\cdot 10^1$ \\
$c_7$      & $(1.73\pm 0.31)\cdot 10^{-1}$ \\
$c_8$      & $(1.98\pm 0.69)\cdot 10^1$ \\
$c_9$      & $(1.60\pm 0.44)\cdot 10^1$ \\
\hline
\end{tabular}

}
\vspace{3ex}
\caption{\label{tab:swpar_values_errors}
  Extended Sanford-Wang parameters and errors obtained by fitting the
  dataset. The errors refer to the 68.27\% confidence
  level for eight parameters ($\Delta\chi^2=9.30$).
}
\end{table*}
\begin{table*}
\centerline{
\begin{tabular}{|c|r r r r r r r r|}
\hline
{\bf Parameter} & $c_1$  & $c_2$  & $c_3$  & $c_4=c_5$  & $c_6$  & $c_7$  & $c_8$ & $c_9$ \\
\hline
$c_1$     &  1.000 &        &        &        &        &        &        &       \\
$c_2$     &  0.341 &  1.000 &        &        &        &        &        &       \\
$c_3$     &  0.099 &  0.562 &  1.000 &        &        &        &        &       \\
$c_4=c_5$ & -0.696 & -0.730 &  0.004 &  1.000 &        &        &        &       \\
$c_6$     & -0.309 &  0.288 &  0.735 &  0.358 &  1.000 &        &        &       \\
$c_7$     & -0.609 &  0.066 & -0.221 &  0.030 &  0.005 &  1.000 &        &       \\
$c_8$     & -0.170 & -0.030 & -0.270 & -0.173 & -0.433 &  0.672 &  1.000 &       \\
$c_9$     & -0.250 &  0.270 &  0.819 &  0.368 &  0.973 & -0.060 & -0.393 & 1.000 \\
\hline
\end{tabular}

}
\vspace{3ex}
\caption{\label{tab:swpar_correlations}
Correlation coefficients among the extended Sanford-Wang parameters, obtained
 by fitting the data.
}
\end{table*}

The HARP cross-section measurement is compared to the best-fit parametrization of Eqs.~\ref{eq:swformula3} and \ref{eq:swformula2}, and Table~\ref{tab:swpar_values_errors}, in Figs.~\ref{fig:xsecBins} and~\ref{fig:xsec1D}. Also by looking at Fig.~\ref{fig:xsecBins}, one can qualitatively conclude that the proposed generalization of the Sanford-Wang parametrization provides a reasonable description of the data spectral features over the entire pion phase space measured. On the other hand, as already noted in \cite{ref:alPaper}, we remark that the goodness-of-fit depends on the correlations among the HARP cross-section uncertainties in different $(p,\theta)$ bins, and therefore cannot be inferred solely from Fig.~\ref{fig:xsecBins}. 

 We defer to a later publication, which should include a more comprehensive study of $\pi^+$ production at various beam momenta and from various nuclear targets, a more complete discussion on the adequacy of parametrization-driven models such as Sanford-Wang (or simple modifications of) to describe HARP hadron production data.

\section{Relevance of HARP beryllium results for neutrino experiments}\label{sec:miniboone}

The Booster neutrino beam at the Fermi National Accelerator Laboratory in Batavia, Illinois is created
from the decay of charged mesons passing through a 50~\m open decay region.  These mesons are produced
when 8.9 \GeVc momentum protons are impinged upon a 71~\cm~long (1.7$\lambda$) 
by 1~\cm~diameter beryllium target located at the upstream end
of a magnetic focusing horn. This neutrino beam has been used by the MiniBooNE experiment
since September, 2002 and will be used by the SciBooNE experiment starting in summer, 2007.

The MiniBooNE (E898) experiment at Fermilab~\cite{ref:minibooneProposal} was designed to 
address the yet unconfirmed oscillation signal reported by the 
LSND collaboration \cite{ref:LSND}.  MiniBooNE has been searching for the appearance of electron neutrinos in a beam
that is predominantly muon flavor with an \emph{L/E} similar to LSND but with substantially differing systematics.
Additionally, the MiniBooNE detector can be used to make neutrino interaction cross-section measurements
for both charged-current and neutral-current processes. An important systematic for these analyses arises 
from the prediction of the fluxes of different neutrino flavors at the MiniBooNE detector.  For the
\nue appearance search the effect of the normalization uncertainty on \piplus production is largely reduced
by a constraint provided by the \emph{in situ} measurement of \numu charged-current quasi-elastic events. 
Neutrino cross-section measurements at MiniBooNE, however, will directly benefit from 
reductions in neutrino flux normalization uncertainties enabled by these hadron data.

The neutrino flux prediction at the MiniBooNE detector is generated
using a Monte Carlo simulation implemented in Geant4 \cite{ref:geant4}. 
Primary meson production rates are presently determined by
fitting the empirical parametrization of Sanford and Wang \cite{ref:SW} to production data in the relevant region. 
The results presented here, being for protons at exactly the Booster beam energy, are a 
critical addition to the global Sanford-Wang parametrization fits.\footnote{Cross-section data from 
the Brookhaven E910 experiment are also used 
to provide an additional constraint to the Sanford-Wang formula at angles larger than 210 mrad.  
These data are from 6.4 and 12.3 \GeVc proton beams on a beryllium target.} 

A flux prediction based on these data has been used by the MiniBooNE 
collaboration in their search for \nue appearance at $\Delta m^2 \sim 1 \hspace{1ex}\mathrm{eV}^2$ \cite{ref:minibooneOsc}. 
The systematic uncertainty on their prediction of \nue from $\pi^+ \rightarrow \mu^+ \rightarrow \nu_e$ as a 
background to the oscillation analysis is 8\% \cite{ref:minibooneOsc} with only a 3\% 
contribution from the \piplus production model based on these 
data.  In contrast, the uncertainty on the prediction of \nue from
kaon decays is 35\% \cite{ref:minibooneOsc}, the largest contribution (25\%) coming from the production model of 
kaons in p+Be interactions, also based on parameterizations of available cross-section data.

%

\begin{figure}
  \begin{center}
    \includegraphics[width=6.5cm]{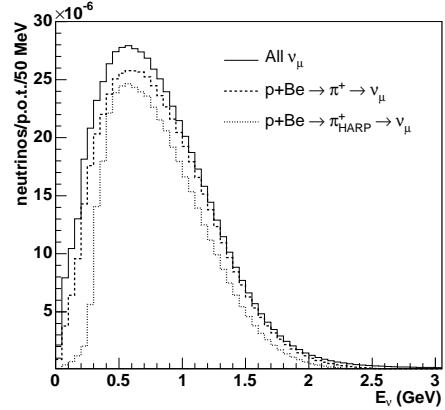}
    \caption{\label{fig:mbFlux} Predicted muon neutrino flux at the MiniBooNE detector from a Geant4
      simulation of the Booster Neutrino Beam at Fermilab based on the parametrization of the HARP \pip production cross-section 
   (Eqs.~\ref{eq:swformula} and \ref{eq:swformula2} and Table~\ref{tab:swpar_values_errors}).  
   The solid curve is the total muon neutrino flux, while the
      dashed curve is the part of the $\nu_{\mu}$ flux coming from the decay of $\pi^{+}$ created in proton-beryllium collisions.
      The primary production of positive pions is based on a parametrization of the HARP $\pi^{+}$ 
      cross-section measurements presented in this paper and represents the predominant source of $\nu_{\mu}$ at MiniBooNE. The dotted
      histogram shows the part of the $\nu_{\mu}$ flux coming from the decay of $\pi^{+}$'s that are within the kinematic boundaries
      of the measurement presented here, $0.75 \ \GeVc \leq p_{\pi} \leq 6.5$ \GeVc and $0.030 \
      \rad \leq \theta_{\pi} \leq 0.210 \ \rad$.}  
    \end{center}
\end{figure}

\begin{figure}
  \begin{center}
    \includegraphics[width=6.5cm]{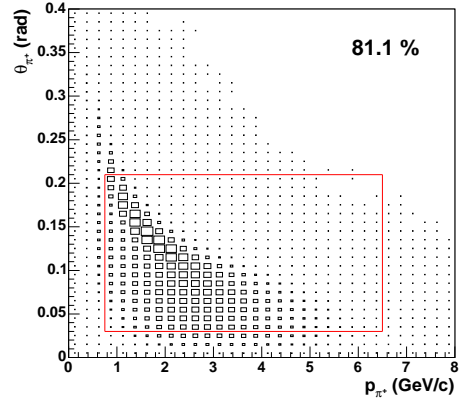}
    \caption{\label{fig:mbPions2D} $p$-$\theta$ distribution of $\pi^{+}$'s which decay to a muon neutrino passing through the MiniBooNE
    detector according to a Monte Carlo simulation.  The box marks the kinematic region of the measurement presented
    here and contains 81.1\% of the pions contributing to the neutrino flux at MiniBooNE. }
  \end{center}
\end{figure}

Using the complete MiniBooNE beam Monte Carlo we can illustrate the direct impact of the HARP data on the MiniBooNE flux predictions.  
The dominant channel leading to a muon neutrino in the detector is 
$\mathrm{p} + \mathrm{Be} \rightarrow \pi^{+} \rightarrow \nu_{\mu}$. 
Fig. \ref{fig:mbFlux} shows the 
total $\nu_{\mu}$ flux (solid) according to the simulation as well as the part coming directly from the sequence listed above (dashed).
The curves are obtained from a Geant4 simulation of the Booster Neutrino Beam at Fermilab 
based on the parametrization of the HARP \pip production cross-section given in 
Eqs.~\ref{eq:swformula} and \ref{eq:swformula2} and Table~\ref{tab:swpar_values_errors}. 
Figure~\ref{fig:mbPions2D} shows the kinematic distribution of $\pi^{+}$'s which result in a $\nu_{\mu}$ in the MiniBooNE detector.  The box 
outlines the kinematic range of the measurements described in this paper.
The simulation indicates that $>$80\% of the relevant pions
come from within this region.  The neutrinos produced by the subset of pion phase-space directly covered by this measurement are
shown by the dotted histogram in Fig. \ref{fig:mbFlux}.  While the coverage of these data is being displayed for the MiniBooNE 
detector, a similar coverage is expected for the SciBooNE detector~\cite{ref:sciboone} located in the same neutrino beam at Fermilab.

\section{Summary and conclusions}\label{sec:conclusions}

In this paper we have presented a measurement of the double-differential production cross-section
of positive pions in the collision of 8.9 \GeVc protons with a beryllium target.  The data have been
reported in 78 bins of pion momentum and angle in the kinematic
range from $0.75 \ \GeVc \leq p_{\pi} \leq 6.5$ \GeVc and $0.030 \
\rad \leq \theta_{\pi} \leq 0.210 \ \rad$.   
A systematic error analysis has been performed yielding an average point-to-point error of 9.8\% 
(statistical + systematic) and
an uncertainty on the total integrated cross-section of 4.9\%.  
Further, the data have been fitted to a modified form of the empirical parameterization
of Sanford and Wang and the resulting parameters provided.    

These production data have direct relevance for the prediction of a \numu flux for MiniBooNE, an
experiment searching for \numu $\rightarrow$ \nue oscillations using the Booster neutrino beam line at
Fermi National Accelerator Laboratory, and SciBooNE, an experiment designed to measure \numu cross-sections
in the 1 \GeV neutrino energy region using the same beam.  Final flux predictions for these experiments will be based on the
results presented here and published elsewhere by the MiniBooNE and SciBooNE collaborations.

\section{Acknowledgments}

We gratefully acknowledge the help and support of the PS beam staff
and of the numerous technical collaborators who contributed to the
detector design, construction, commissioning and operation.  
In particular, we would like to thank 
G.~Barichello, 
R.~Brocard, 
K.~Burin, 
V.~Carassiti, 
F.~Chignoli, 
D.~Conventi, 
G.~Decreuse, 
M.~Delattre, 
C.~Detraz,   
A.~Domeniconi, 
M.~Dwuznik,    
F.~Evangelisti, 
B.~Friend, 
A.~Iaciofano,  
I.~Krasin,  
D.~Lacroix, 
J.-C.~Legrand, 
M.~Lobello,  
M.~Lollo, 
J.~Loquet, 
F.~Marinilli, 
J.~Mulon, 
L.~Musa, 
R.~Nicholson, 
A.~Pepato, 
P.~Petev,  
X.~Pons, 
I.~Rusinov, 
M.~Scandurra, 
E.~Usenko, 
and 
R.~van der Vlugt, 
for their support in the construction of the detector.
The collaboration acknowledges the major contributions and advice of
M.~Baldo-Ceolin, 
L.~Linssen, 
M.T.~Muciaccia and A. Pullia
during the construction of the experiment.
The collaboration is indebted to 
V.~Ableev,
F.~Bergsma,
P.~Binko,
E.~Boter,
M.~Calvi, 
C.~Cavion, 
A.~Chukanov,  
M.~Doucet,
D.~D\"{u}llmann,
V.~Ermilova, 
W.~Flegel,
Y.~Hayato,
A.~Ichikawa,
A.~Ivanchenko,
O.~Klimov,
T.~Kobayashi,
D.~Kustov, 
M.~Laveder, 
M.~Mass,
H.~Meinhard,
A.~Menegolli, 
T.~Nakaya,
K.~Nishikawa,
M.~Pasquali,
M.~Placentino,
S.~Simone,
S.~Troquereau,
S.~Ueda and A.~Valassi
for their contributions to the experiment.

We acknowledge the contributions of 
F.~Dydak and
J.~Wotschack 
to the work described in this paper.

We are indebted to the MiniBooNE collaboration who made available their
beam-line simulation for the calculation of the predicted
neutrino fluxes at their detector.

 The experiment was made possible by grants from
the Institut Interuniversitaire des Sciences Nucl\'eair\-es and the
Interuniversitair Instituut voor Kernwetenschappen (Belgium), 
Ministerio de Educacion y Ciencia, Grant FPA2003-06921-c02-02 and
Generalitat Valenciana, grant GV00-054-1,
CERN (Geneva, Switzerland), 
the German Bundesministerium f\"ur Bildung und Forschung (Germany), 
the Istituto Na\-zio\-na\-le di Fisica Nucleare (Italy), 
INR RAS (Moscow) and the Particle Physics and Astronomy Research Council (UK).
We gratefully acknowledge their support.


\newpage

\begin{appendix}

\label{app:data}

\begin{table}[!h]
  \centerline{
    \begin{tabular}{|c|c|c|c|rcr|} \hline
$\theta_{\hbox{\small min}}$ &
$\theta_{\hbox{\small max}}$ &
$p_{\hbox{\small min}}$ &
$p_{\hbox{\small max}}$ &
\multicolumn{3}{c|}{$d^2\sigma^{\pi^+}/(dpd\Omega)$} 
\\
(mrad) & (mrad) & (GeV/c) & (GeV/c) &
\multicolumn{3}{c|}{(mb/(GeV/c sr))}
\\ \hline
 30 &  60 & 0.75 & 1.00 & $174.7$ & $\pm$ & $23.6$ \\
    &     & 1.00 & 1.25 & $174.8$ & $\pm$ & $24.4$ \\
    &     & 1.25 & 1.50 & $187.2$ & $\pm$ & $21.7$ \\
    &     & 1.50 & 1.75 & $180.2$ & $\pm$ & $22.5$ \\
    &     & 1.75 & 2.00 & $200.3$ & $\pm$ & $18.9$ \\
    &     & 2.00 & 2.25 & $210.5$ & $\pm$ & $17.8$ \\
    &     & 2.25 & 2.50 & $190.4$ & $\pm$ & $15.5$ \\
    &     & 2.50 & 2.75 & $180.2$ & $\pm$ & $14.7$ \\
    &     & 2.75 & 3.00 & $168.4$ & $\pm$ & $14.4$ \\
    &     & 3.00 & 3.25 & $125.3$ & $\pm$ & $10.5$ \\
    &     & 3.25 & 4.00 & $ 99.1$ & $\pm$ & $ 8.3$ \\
    &     & 4.00 & 5.00 & $ 64.4$ & $\pm$ & $ 6.9$ \\
    &     & 5.00 & 6.50 & $ 29.2$ & $\pm$ & $ 5.9$ \\ \hline
 60 &  90 & 0.75 & 1.00 & $150.1$ & $\pm$ & $17.2$ \\
    &     & 1.00 & 1.25 & $161.1$ & $\pm$ & $15.5$ \\
    &     & 1.25 & 1.50 & $167.8$ & $\pm$ & $16.7$ \\
    &     & 1.50 & 1.75 & $174.6$ & $\pm$ & $17.0$ \\
    &     & 1.75 & 2.00 & $184.1$ & $\pm$ & $15.0$ \\
    &     & 2.00 & 2.25 & $186.2$ & $\pm$ & $12.1$ \\
    &     & 2.25 & 2.50 & $177.4$ & $\pm$ & $11.1$ \\
    &     & 2.50 & 2.75 & $149.7$ & $\pm$ & $ 9.5$ \\
    &     & 2.75 & 3.00 & $123.6$ & $\pm$ & $ 8.3$ \\
    &     & 3.00 & 3.25 & $ 92.0$ & $\pm$ & $ 8.0$ \\
    &     & 3.25 & 4.00 & $ 68.8$ & $\pm$ & $ 5.4$ \\
    &     & 4.00 & 5.00 & $ 43.2$ & $\pm$ & $ 3.9$ \\
    &     & 5.00 & 6.50 & $ 17.5$ & $\pm$ & $ 2.8$ \\ \hline
 90 & 120 & 0.75 & 1.00 & $180.7$ & $\pm$ & $17.3$ \\
    &     & 1.00 & 1.25 & $178.8$ & $\pm$ & $18.4$ \\
    &     & 1.25 & 1.50 & $180.2$ & $\pm$ & $13.1$ \\
    &     & 1.50 & 1.75 & $146.1$ & $\pm$ & $13.0$ \\
    &     & 1.75 & 2.00 & $142.2$ & $\pm$ & $12.8$ \\
    &     & 2.00 & 2.25 & $126.9$ & $\pm$ & $10.3$ \\
    &     & 2.25 & 2.50 & $118.8$ & $\pm$ & $ 8.5$ \\
    &     & 2.50 & 2.75 & $ 95.2$ & $\pm$ & $ 7.0$ \\
    &     & 2.75 & 3.00 & $ 73.8$ & $\pm$ & $ 6.8$ \\
    &     & 3.00 & 3.25 & $ 57.9$ & $\pm$ & $ 5.5$ \\
    &     & 3.25 & 4.00 & $ 43.7$ & $\pm$ & $ 3.3$ \\
    &     & 4.00 & 5.00 & $ 22.9$ & $\pm$ & $ 2.8$ \\
    &     & 5.00 & 6.50 & $  8.3$ & $\pm$ & $ 1.5$ \\ \hline
\end{tabular}

  }
\end{table}

\begin{table}
  \centerline{
    \begin{tabular}{|c|c|c|c|rcr|} \hline
$\theta_{\hbox{\small min}}$ &
$\theta_{\hbox{\small max}}$ &
$p_{\hbox{\small min}}$ &
$p_{\hbox{\small max}}$ &
\multicolumn{3}{c|}{$d^2\sigma^{\pi^+}/(dpd\Omega)$} 
\\
(mrad) & (mrad) & (GeV/c) & (GeV/c) &
\multicolumn{3}{c|}{(mb/(GeV/c sr))}
\\ \hline
120 & 150 & 0.75 & 1.00 & $168.7$ & $\pm$ & $16.9$ \\
    &     & 1.00 & 1.25 & $155.3$ & $\pm$ & $16.5$ \\
    &     & 1.25 & 1.50 & $153.7$ & $\pm$ & $13.2$ \\
    &     & 1.50 & 1.75 & $142.9$ & $\pm$ & $11.1$ \\
    &     & 1.75 & 2.00 & $118.4$ & $\pm$ & $ 9.6$ \\
    &     & 2.00 & 2.25 & $ 97.1$ & $\pm$ & $ 8.4$ \\
    &     & 2.25 & 2.50 & $ 86.0$ & $\pm$ & $ 6.6$ \\
    &     & 2.50 & 2.75 & $ 66.4$ & $\pm$ & $ 6.5$ \\
    &     & 2.75 & 3.00 & $ 54.6$ & $\pm$ & $ 5.5$ \\
    &     & 3.00 & 3.25 & $ 45.2$ & $\pm$ & $ 4.8$ \\
    &     & 3.25 & 4.00 & $ 27.6$ & $\pm$ & $ 3.2$ \\
    &     & 4.00 & 5.00 & $ 13.5$ & $\pm$ & $ 2.2$ \\
    &     & 5.00 & 6.50 & $  4.9$ & $\pm$ & $ 0.8$ \\ \hline
150 & 180 & 0.75 & 1.00 & $165.2$ & $\pm$ & $20.3$ \\
    &     & 1.00 & 1.25 & $173.9$ & $\pm$ & $15.7$ \\
    &     & 1.25 & 1.50 & $143.8$ & $\pm$ & $15.4$ \\
    &     & 1.50 & 1.75 & $116.1$ & $\pm$ & $10.8$ \\
    &     & 1.75 & 2.00 & $ 97.4$ & $\pm$ & $ 8.6$ \\
    &     & 2.00 & 2.25 & $ 68.2$ & $\pm$ & $ 6.1$ \\
    &     & 2.25 & 2.50 & $ 52.1$ & $\pm$ & $ 5.6$ \\
    &     & 2.50 & 2.75 & $ 41.5$ & $\pm$ & $ 3.8$ \\
    &     & 2.75 & 3.00 & $ 35.0$ & $\pm$ & $ 3.8$ \\
    &     & 3.00 & 3.25 & $ 24.8$ & $\pm$ & $ 3.0$ \\
    &     & 3.25 & 4.00 & $ 17.4$ & $\pm$ & $ 2.1$ \\
    &     & 4.00 & 5.00 & $  7.8$ & $\pm$ & $ 1.2$ \\
    &     & 5.00 & 6.50 & $  2.8$ & $\pm$ & $ 0.3$ \\ \hline
180 & 210 & 0.75 & 1.00 & $171.3$ & $\pm$ & $29.8$ \\
    &     & 1.00 & 1.25 & $131.9$ & $\pm$ & $15.0$ \\
    &     & 1.25 & 1.50 & $101.6$ & $\pm$ & $12.3$ \\
    &     & 1.50 & 1.75 & $ 63.5$ & $\pm$ & $ 8.9$ \\
    &     & 1.75 & 2.00 & $ 49.2$ & $\pm$ & $ 6.7$ \\
    &     & 2.00 & 2.25 & $ 39.7$ & $\pm$ & $ 5.1$ \\
    &     & 2.25 & 2.50 & $ 28.1$ & $\pm$ & $ 4.2$ \\
    &     & 2.50 & 2.75 & $ 21.8$ & $\pm$ & $ 3.0$ \\
    &     & 2.75 & 3.00 & $ 15.6$ & $\pm$ & $ 2.2$ \\
    &     & 3.00 & 3.25 & $ 13.1$ & $\pm$ & $ 1.8$ \\
    &     & 3.25 & 4.00 & $  9.9$ & $\pm$ & $ 1.3$ \\
    &     & 4.00 & 5.00 & $  5.6$ & $\pm$ & $ 0.8$ \\
    &     & 5.00 & 6.50 & $  1.8$ & $\pm$ & $ 0.4$ \\ \hline
\end{tabular}

  }
  \vspace{2ex}
  \caption{\label{tab:xsec_results2}
    HARP results for the double-differential $\pi^+$ production
    cross-section in the laboratory system,
    $d^2\sigma^{\pi^+}/(dpd\Omega)$. Each row refers to a
    different $(p_{\hbox{\small min}} \le p<p_{\hbox{\small max}},
    \theta_{\hbox{\small min}} \le \theta<\theta_{\hbox{\small max}})$ bin,
    where $p$ and $\theta$ are the pion momentum and polar angle, respectively.
    The central value as well as the square-root of the diagonal elements
    of the covariance matrix are given.}
\end{table}

\end{appendix}

\end{document}